\documentclass[noshowpacs,preprint,aps,pre]{revtex4}
\usepackage[dvips]{graphicx}
\usepackage{latexsym}
\usepackage{epsfig}
\usepackage{bm}
\usepackage{times,psfrag,subfigure}        
\begin{document}

\title{Modelling contact angle hysteresis on chemically patterned and superhydrophobic surfaces}
\author{H. Kusumaatmaja}
\author{J. M. Yeomans}
\affiliation{The Rudolf Peierls Centre for Theoretical Physics, Oxford University, 1 Keble Road, Oxford OX1 3NP, U.K.}
\date{\today}

\begin{abstract}

We investigate contact angle hysteresis on chemically patterned and 
superhydrophobic surfaces, as the drop volume is quasi-statically increased and 
decreased. We consider both two, and three, dimensions using analytical and numerical 
approaches to minimise the free energy of the drop. In two dimensions we find, in 
agreement with other authors, a slip, jump, stick motion of the contact line. In three 
dimensions this behaviour persists, but the position and magnitude of the contact line 
jumps are sensitive to the details of the surface patterning. In two dimensions we 
identify analytically the advancing and receding contact angles on the different surfaces 
and we use numerical insights to argue that these provide bounds for the three
dimensional cases. We present explicit simulations to show that a simple average
over the disorder is not sufficient to predict the details of the contact angle
hysteresis, and to support an explanation for the low contact angle hysteresis of
suspended drops on superhydrophobic surfaces. 

\end{abstract}
\maketitle


\section{Introduction}

Our aim in this paper is to investigate contact angle hysteresis on chemically patterned 
and superhydrophobic surfaces. 
For a surface that is flat and chemically homogeneous, a liquid drop will form a 
spherical cap with a contact angle $\theta$ given by Young's law \cite{Young}
\begin{equation}
\cos \theta = \frac{\sigma_{GS} - \sigma_{LS}}{\sigma_{LG}} \, , \label{eq1}
\end{equation}
where $\sigma_{GS}$, $\sigma_{LS}$ and $\sigma_{LG}$ and the gas--solid, liquid--solid and 
liquid--gas surface tensions respectively. This shape corresponds to the minimum free 
energy configuration of the system, assuming that the drop size is much smaller than the 
capillary length so that gravity can be neglected.

Real surfaces are, however, neither perfectly flat nor chemically homogeneous. These 
inhomogeneities result in the existence of multiple local free energy minima, not
just the global minimum prescribed by Young's formula (\ref{eq1}). This can cause pinning 
of the contact line and lead to drop shapes which depend not only on the thermodynamic 
variables describing the state of the drop, but also on the path by which that state was 
achieved. This phenomenon is known as contact angle hysteresis.

There are several different manifestations of contact angle hysteresis commonly reported in the 
literature and it is important to distinguish clearly between them. One favoured 
experimental approach to measure the contact angle of a drop is to slowly increase the 
volume until the drop starts to spread. The angle at which spreading occurs, which 
corresponds to the drop being able to cross any free energy barrier which impedes its 
motion, is termed the advancing contact angle. Similarly, as the drop volume is 
quasi-statically decreased, the contact line first moves at the receding contact angle. 
The difference between the advancing and receding contact angles is the contact angle 
hysteresis, zero for a perfect substrate, but possibly $10^{\mathrm{o}}$ or more, and notoriously 
difficult to measure, for a real surface. We shall concentrate on this set-up here and 
shall term it unforced, static hysteresis.

It is also possible to push a drop across a surface and measure the advancing angle, at 
the front of the drop, and the receding contact angle, at the rear, when it first moves.
We shall refer to this case as forced, static hysteresis and comment on it further in the 
discussion. However it is important to point out that the measured unforced and forced 
contact angle hystersis for a given drop on a given surface need not necessarily be the 
same as a forced drop will be deformed and the free energy barriers will depend on the 
shape of the drop.

In the literature there is also mention of dynamic contact angle hysteresis, the 
difference in angles at the front and rear of a moving drop. This is again a different 
situation, which cannot be treated by equilibrium statistical mechanics, and we shall not 
consider it further here.

Understanding contact angle hysteresis quantitatively is not easy because it depends on 
the details of the surface inhomogeneities which will, in general, be random in position 
and size. However recent advances have allowed surfaces to be fabricated with 
well-defined chemical patterning -- areas of differing contact angle -- which can be 
varied in size relative to the size of a drop (e.g. \cite{Lenz1,Darhuber1,Leopoldes1}). 
This is allowing more quantitative measurements of how advancing and receding contact angles, 
and contact angle hysteresis, depend on the details of the surface patterning.

Another exciting development is the fabrication of superhydrophobic surfaces \cite{Quere1,Quere2,McCarthy1}. When 
surfaces with an intrinsic hydrophobic contact angle are covered with micron scale posts, 
the macroscopic contact angle increases to, in certain cases, close to $180^{\mathrm{o}}$. 
Drops can either lie in a suspended, or Cassie-Baxter \cite{Cassie}, state on top of the 
posts or a collapsed, or Wenzel \cite{Wenzel}, state filling the interstices between them. 
Moreover on these surfaces, drops can roll very easily \cite{Quere1,Quere2,McCarthy1,McCarthy2,McCarthy3} and contact angle hysteresis 
is important in understanding why. It is of interest to ask how the position, size and spacing of the 
posts affects the hysteresis, questions which have caused some controversy in the literature
\cite{McCarthy1,McCarthy2,McCarthy3,McHale1,Extrand}, for example whether contact angle hysteresis is primarily a 
surface (e.g. \cite{McHale1}), or a contact line (e.g. \cite{McCarthy1,McCarthy2,McCarthy3,Extrand}), effect. Very recent 
experiments by Dorrer {\em et al.} \cite{Dorrer} show that the advancing contact angle is 
constant, while the receding contact angle varies as a function of the post size and spacing. 

The role of free energy barriers in contact angle hysteresis has been known for some time 
\cite{Derjaguin,deGennes2,Iwamatsu1,Johnson1,Huh1,Johnson3,Marmur1,Marmur3,Huh2,Amirfazli,Schwartz1,Schwartz2,deGennes1,
Marmur2,Chatain1,Kong}. Several authors have considered the two dimensional (or axisymmetric three 
dimensional) problem of a drop moving over chemically striped surfaces \cite{Iwamatsu1,Johnson1}, 
sinusoidal surfaces \cite{Huh1,Johnson3,Marmur1,Marmur3}, or surfaces with posts \cite{Huh2,Amirfazli} 
using analytic or numerical techniques to minimise the free energy. Extensions to three dimensions 
have also been attempted \cite{Johnson1,deGennes1,Marmur2,Chatain1,Kong,Schwartz1,Schwartz2}. 
These efforts include surfaces with random isolated heterogeneities \cite{deGennes1}, 
striped patterns \cite{Marmur2}, and regular lattices of posts \cite{Chatain1,Kong}
or chemical patches \cite{Johnson1,Schwartz1,Schwartz2}. The common themes in all these works are the 
pinning of the contact line, the existence of multiple local energy minima, and the occurence 
of a slip-jump-stick behaviour. 

In this paper we investigate contact angle hysteresis on chemically patterned and 
superhydrophobic surfaces, as the drop volume is quasi-statically increased and 
decreased. We consider both two, and three, dimensions using analytical and numerical 
approaches to minimise the free energy of the drop. In two dimensions, on a surface 
striped with regions of different equilibrium contact angle, $\theta_e$, we find, in 
agreement with other authors \cite{Iwamatsu1,Johnson1}, a slip, jump, stick motion of 
the contact line. The advancing and receding contact angles are equal to the maximum 
and minimum values of the $\theta_e$ respectively. In three dimensions these values 
provide bounds, but the contact angle hysteresis is reduced by the free energy associated 
with surface distortion \cite{Schwartz1,Schwartz2,deGennes1}. A stick, slip, jump behaviour persists,
but we caution that the definition of a single macroscopic contact angle is problematic 
for patterns of order the drop size. The position and magnitude of the contact line jumps 
are sensitive to the details of the surface patterning and can be different in different 
directions relative to that patterning.

For drops suspended on superhydrophobic surfaces we find that, in two dimensions, the 
advancing contact angle is (ideally) $180^{\mathrm{o}}$, in agreement with \cite{Huh2} 
and the receding angle is $\theta_e$, the intrinsic contact angle of the surface. For 
collapsed drops in two dimensions the advancing contact angle is also $180^{\mathrm{o}}$ 
but the receding angle is $\theta_e - 90^{\mathrm{o}}$ because the contact line has to dewet 
the sides of the posts. In three dimensions, for both suspended and collapsed drops, the 
advancing angle remains close to $180^{\mathrm{o}}$ \cite{McCarthy1,McCarthy2,McCarthy3,Dorrer,Chatain1}, but the 
receding contact angle is increased. However, the receding angle still remains considerably smaller for the 
collapsed state and the contact line pinning is much stonger for the collapsed state. 
As a result, the hysteresis of suspended drops is, in general, much smaller than that of 
collapsed drops on the same surface. Again we argue that a simple average over the disorder 
is not sufficient to predict the details of the contact angle hysteresis. 

We organise the paper as follows: in the next section we list the equations describing 
the dynamics and thermodynamics of the drops we consider. We briefly describe the lattice 
Boltzmann algorithm used to solve these equations and, in particular, the numerical scheme 
employed to incrementally add or remove the drop volume. In sections III and IV, we focus on
chemically patterned surfaces. We then discuss topologically patterned surfaces in 
sections V--VII treating first the suspended, and then the collapsed, state. In each case 
we consider two, and then three, dimensions. Section IX summarises our conclusions and 
discusses areas for future research. 


\section{The model drop}

We choose to model the equilibrium properties of the drop by a continuum free energy \cite{Briant1}
\begin{equation} 
\Psi = \int_V (\psi_b(n)+\frac{\kappa}{2} (\partial_{\alpha}n)^2) dV
+ \int_S \psi_s(n_s) dS . \label{eq3}
\end{equation}
$\psi_b(n)$ is a bulk free energy term which we take to be \cite{Briant1}
\begin{equation}
\psi_b (n) = p_c (\nu_n+1)^2 (\nu_n^2-2\nu_n+3-2\beta\tau_w) \, ,
\end{equation}
where $\nu_n = {(n-n_c)}/{n_c}$, $\tau_w = {(T_c-T)}/{T_c}$ and $n$, $n_c$, $T$, $T_c$ and $p_c$ 
are the local density, critical density, local temperature, critical temperature and critical 
pressure of the fluid respectively. This choice of free energy leads to two coexisting bulk phases 
of density $n_c(1\pm\sqrt{\beta\tau_w})$, which represent the liquid drop and surrounding gas 
respectively. Varying $\beta$ has the effects of varying the densities, surface tension, and 
interface width; we typically choose $\beta = 0.1$. 

The second term in Eq.\ (\ref{eq3}) models the free energy associated with any interfaces in the system. $\kappa$ is related to the 
liquid--gas surface tension and interface width via $\sigma_{lg} = {(4\sqrt{2\kappa p_c} (\beta\tau_w)^{3/2} n_c)}/3$
and $\xi = (\kappa n_c^2/4\beta\tau_w p_c)^{1/2}$ \cite{Briant1}. Unless stated otherwise, we use $\kappa = 0.004$, 
$p_c = 1/8$, $\tau_w = 0.3$, and $n_c = 3.5$.

The last term in Eq.\ (\ref{eq3}) describes the interactions between the fluid and the solid surface. 
Following Cahn \cite{Cahn} the surface energy density is taken to be $\psi_s (n) = -\phi \, n_s$,
where $n_s$ is the value of the fluid density at the surface.
The strength of interaction, and hence the local equilibrium contact angle, is parameterised 
by the variable $\phi$. Minimising the free energy (\ref{eq3}) leads to the boundary 
condition at the surface, 
\begin{equation}
\partial_{\perp}n = -\phi/\kappa \, , \label{dn}
\end{equation}
and a relation between $\phi$ and the equilibrium contact angle $\theta_e$ \cite{Briant1}
\begin{equation}
\phi = 2\beta\tau_w\sqrt{2p_c\kappa} \,\, 
\mathrm{sign}(\frac{\pi}{2}-\theta_e)\sqrt{\cos{\frac{\alpha}{3}}(1-\cos{\frac{\alpha}{3}})} \, , 
\label{eq6}
\end{equation}
where $\alpha=\cos^{-1}{(\sin^2{\theta_e})}$ and the function sign returns the sign of its argument. 
On chemically heterogeneous surfaces the contact angle $\theta_e$ will be a function of position. 
This can be modelled easily within a simulation by using Eq. (\ref{eq6}) to calculate the appropriate 
value of $\phi$ and then constraining the normal 
derivative $\partial_{\perp}n$ at different regions of the surface to take the appropriate values, 
given by Eq. (\ref{dn}). Similar boundary conditions can be used for
surfaces that are not flat: a way to treat the corners and ridges needed
to model superhydrophobic surfaces is described in \cite{Dupuis2}.

The equations of motion of the drop are the continuity and the Navier-Stokes equations
\begin{eqnarray}
&\partial_{t}n+\partial_{\alpha}(nu_{\alpha})=0 \, , \label{eq4}\\
&\partial_{t}(nu_{\alpha})+\partial_{\beta}(nu_{\alpha}u_{\beta}) = 
- \partial_{\beta}P_{\alpha\beta}  
+ \nu \partial_{\beta}[n(\partial_{\beta}u_{\alpha} + \partial_{\alpha}u_{\beta} + 
\delta_{\alpha\beta} \partial_{\gamma} u_{\gamma}) ] + na_{\alpha} \, , \label{eq5}
\end{eqnarray}
where $\mathbf{u}$, $\mathbf{P}$, $\nu$, and $\mathbf{a}$ are the local velocity, pressure tensor, 
kinematic viscosity, and acceleration respectively. The thermodynamic
properties of the drop appear in the equations of motion through the
pressure tensor $\mathbf{P}$ which can be calculated from the free energy \cite{Briant1,Dupuis2}
\begin{eqnarray} 
&P_{\alpha\beta} = (p_{\mathrm{b}}-\frac{\kappa}{2} (\partial_{\alpha}n)^2 -
\kappa n \partial_{\gamma\gamma}n)\delta_{\alpha\beta} 
+ \kappa (\partial_{\alpha}n)(\partial_{\beta}n),  \\ 
&p_{\mathrm{b}} = p_c (\nu_n+1)^2 (3\nu_n^2-2\nu_n+1-2\beta\tau_w) .
\end{eqnarray}
When the drop is at rest $\partial_{\alpha}P_{\alpha\beta} = 0$ and
the free energy (\ref{eq3}) is minimised .

We use a lattice Boltzmann algorithm to solve Eqs. (\ref{eq4})
and (\ref{eq5}). No-slip boundary conditions on the velocity are
imposed on the surfaces adjacent to and opposite the drop and 
periodic boundary conditions are used in the two perpendicular directions.
Details of the lattice Boltzmann approach and of its application to
drop dynamics are given in \cite{Briant1,Dupuis2,Leopoldes1,Succi1,Yeomans}. 

To implement unforced static hysteresis we need to slowly increase or decrease
the drop volume. To do this we vary the drop liquid density by $\pm
0.1\%$. This in turn affects the drop volume as the system
relaxes back to its coexisting equilibrium densities.
Forced hysteresis can be addressed by imposing a body force $na_{\alpha}$
in the Navier-Stokes equation (\ref{eq5}). 


\section{Two dimensional drop on a chemically patterned surface}
\label{chem2D}

In this section we shall investigate a two-dimensional situation, the hysteresis
of a cylindrical drop spreading over a chemically striped surface, where an analytic solution
is possible. We shall demonstrate that the contact line shows a stick, jump, slip
behaviour and that the advancing and receding contact angles are equal to the maximum
and minimum equilibrium contact angles of the surface respectively. We also show that
lattice Boltzmann simulations reproduce the analytic results well.

\begin{figure} 
\begin{center}
\includegraphics[scale=0.5,angle=0]{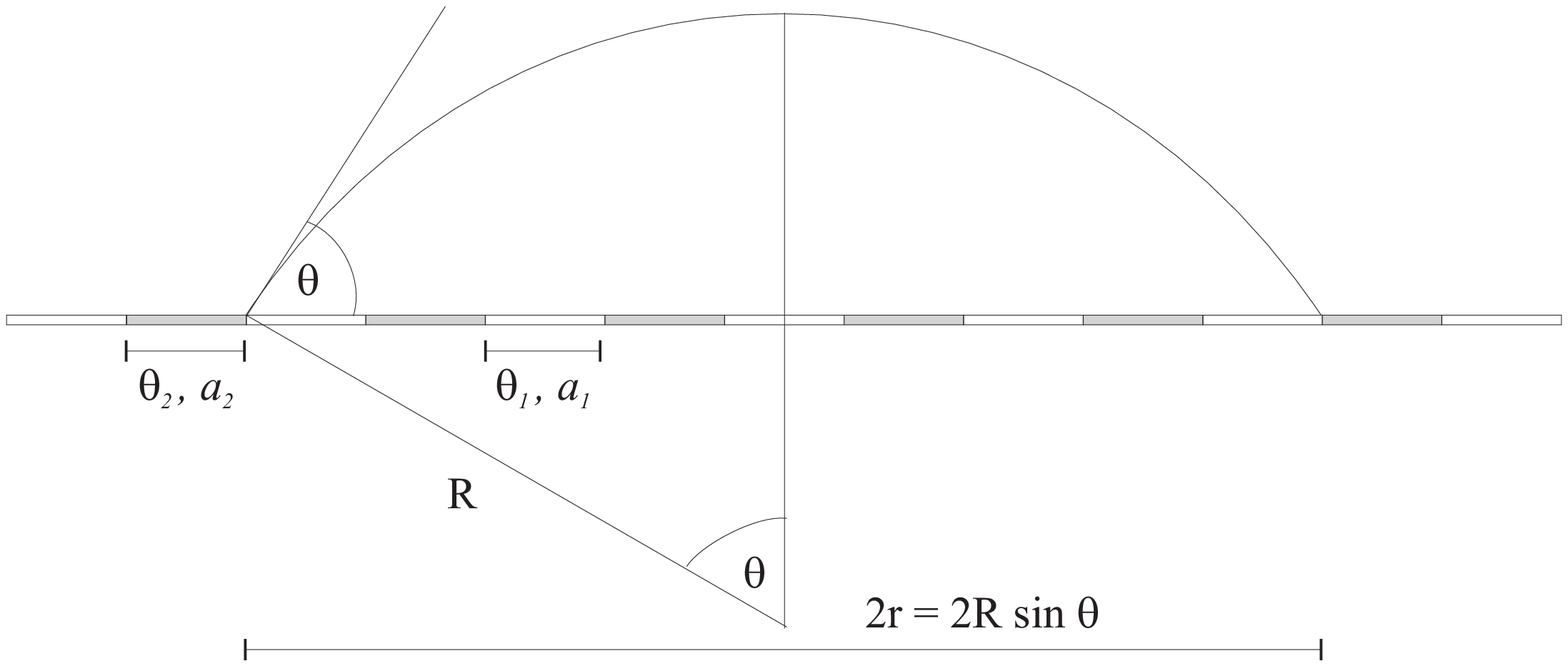}
\caption{
} \label{fig1}
\end{center}
\end{figure}

Consider a cylindrical drop with a {\em macroscopic} contact angle $\theta$ that
forms a spherical cap of radius $R$ as shown in Fig.~\ref{fig1}. (Note
that we reserve $\theta_e$, $\theta_A$, and $\theta_R$ to describe the {\em equilibrium}, 
{\em advancing}, and {\em receding} contact angles on a given surface.) 
The base length of the drop is $2r = 2R \sin{\theta}$. The drop volume per unit
length is
\begin{equation}
S = r^2 \frac{\theta-\sin{\theta}\cos{\theta}}{\sin^2{\theta}} \, \label{eq7}
\end{equation} 
and therefore, for a drop at constant volume,
\begin{equation}
r \, \frac{\sin{\theta}-\theta\cos{\theta}}{\sin^2{\theta}}\,d\theta 
= \, \frac{\sin{\theta}\cos{\theta}-\theta}{\sin{\theta}}\,dr \, .\label{eq14}
\end{equation}
The liquid--gas surface area per unit length is
\begin{equation}
L_{LG} = \frac{2r\theta}{\sin{\theta}}.
\end{equation}

The important contributions to the free energy per unit length of the cylindrical drop
come from the interfacial terms:
\begin{equation}
F = \sigma_{LG} L_{LG} + \int (\sigma_{LS} - \sigma_{GS}) dx \, ,
\label{eq7.5}
\end{equation}
where the integral is taken over the substrate.
Using Young's equation (\ref{eq1}) the reduced free energy follows as
\begin{equation}
f = F/\sigma_{LG} =  \frac{2r\theta}{\sin{\theta}} - \int\cos{\theta_e(x)} dx.  
\label{eq8.5}
\end{equation}

We consider, as in \cite{Johnson3,Iwamatsu1}, the simple case where a drop is 
lying on a surface regularly patterned with (relatively) hydrophilic stripes 
with equilibrium contact angle $\theta_e = \theta_1$ and width $a_1$ and 
(relatively) hydrophobic stripes with parameters $\theta_e = \theta_2$ and width $a_2$. 
Let us further assume that the contact line is initially lying on a hydrophilic
stripe. Our aim is to discuss the motion of the contact line as the drop volume 
slowly increases. The drop initially lies at an angle $\theta=\theta_1$ with 
respect to the solid surface. As the volume is increased the contact line
moves in order to keep the contact angle at a value $\theta=\theta_1$. This
remains true until the contact line reaches the chemical 
border between stripes. It will then be pinned as it costs free energy for the drop to spread
onto the more hydrophobic stripe. The contact line will continue to be
pinned until the free energy cost of distorting the 
liquid--gas interface exceeds the surface energy gain. We demonstrate
that, as expected, this occurs when  $\theta = \theta_2$ \cite{Iwamatsu1,Johnson1}.

From Eq. (\ref{eq8.5}) we may write the reduced free energy of the drop as
\begin{equation}
f = F/\sigma_{LG} =  \frac{2r\theta}{\sin{\theta}} -  (2k+1) \, a_1
  \, \cos{\theta_1} + 2k \, a_2 \, \cos{\theta_2} + 2x \, a_2 \, \cos{\theta_2} 
\end{equation}
where $k$ is an integer and $0 < x < 1$.
Moving the drop a small distance $dr = a_2 \, dx$ onto the hydrophobic stripe the
free energy changes by
\begin{equation}
df = 2\,r\,\frac{\sin{\theta}-\theta\,\cos{\theta}} {\sin^2{\theta}}\,d\theta + 
\frac{2\,\theta}{\sin{\theta}} \, dr - 2\,\cos{\theta_2}\,dr
\end{equation}
which, using the constant volume constraint (Eq. (\ref{eq14})) can be rewritten
\begin{equation} 
df = 2(\cos{\theta}-\cos{\theta_2}) \, dr \, . \label{eq15}
\end{equation}
Hence $df/dr < 0$ and the drop depins when $\theta > \theta_2$.

Increasing the volume further the drop contact line moves with a
constant contact angle $\theta=\theta_2$. Once the drop reaches the hydrophobic--hydrophilic
chemical border it can lower its free energy by immediately spreading
onto the hydrophilic stripe. It will either move to a position where
the contact angle has returned to $\theta_1$ or to the next
hydrophilic--hydrophobic boundary, whichever is reached first. In the 
particular examples we show below, the latter is the case and the contact line 
becomes pinned at the next hydrophilic-hydrophobic
border, with a contact angle $\theta_2 < \theta < \theta_1$.

\begin{figure} 
\begin{center}
\includegraphics[scale=0.8,angle=0]{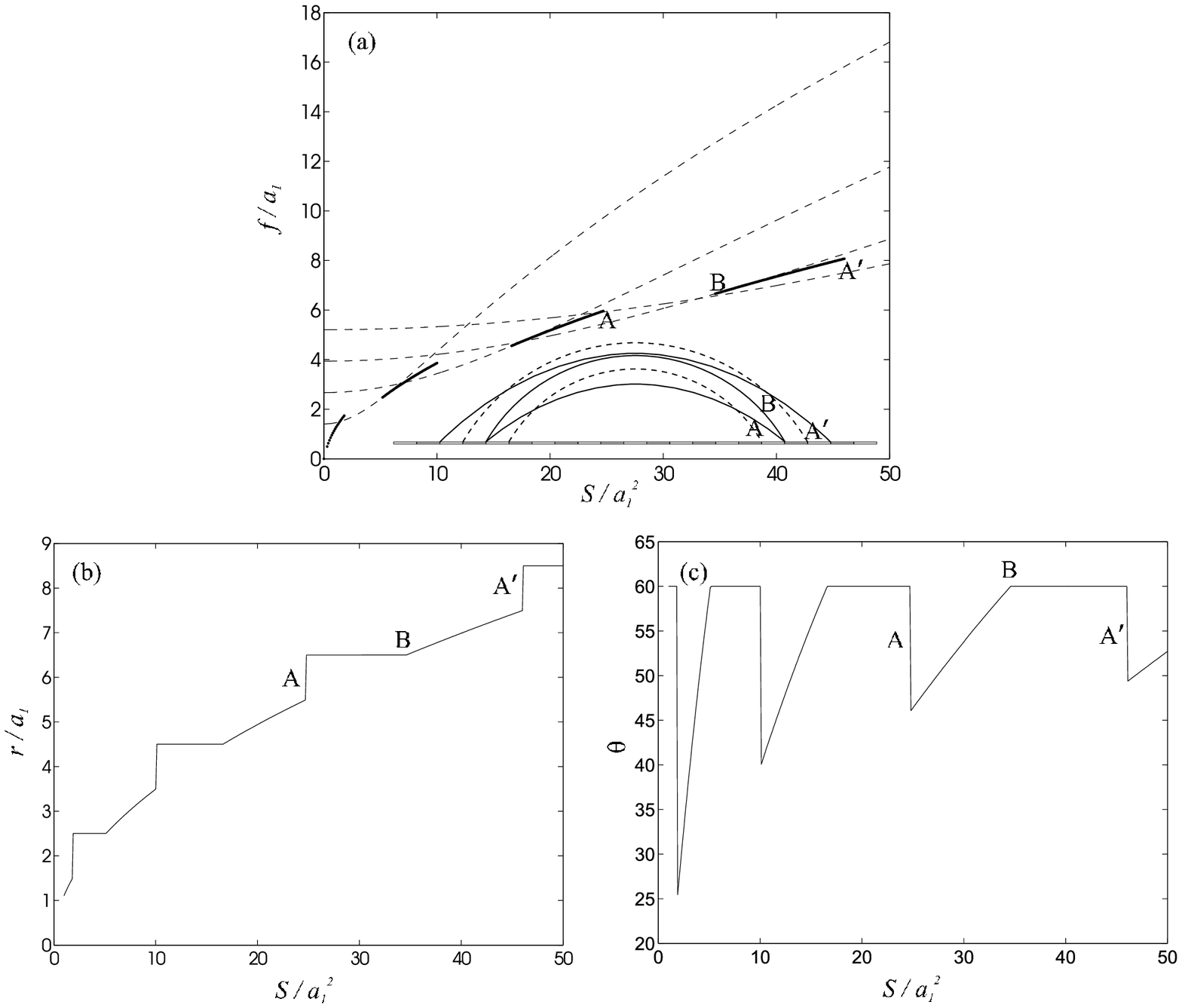}
\caption{
} \label{fig3}
\end{center}
\end{figure}

The free energy (\ref{eq8.5}) of the advancing drop is illustrated in
Fig.~\ref{fig3}(a) for $a_1 / a_2 = 1$, $\theta_1=30^{\mathrm{o}}$ and 
$\theta_2=60^{\mathrm{o}}$. The dashed lines show the free energy of
the drop pinned at successive hydrophilic--hydrophobic boundaries as the
drop volume is increased. The solid line shows the path in free energy space 
taken by the drop as it spreads with a contact angle $\theta=\theta_2$. At A 
the drop reaches, and immediately wets the hydrophilic stripe corresponding to 
a sudden fall in free energy. It reaches the hydrophobic boundary and remains
pinned along AB. At B the contact angle attains the value $\theta_2$ and the
drop spreads across the hydrophobic stripe, reaching the next
hydrophilic stripe at A$^{\prime}$. The solid lines in the inset in Fig.~\ref{fig3}(a) show
the drop profiles at A, B and A$^{\prime}$. The dashed lines in the inset correspond to the drop
profiles at A and A$^{\prime}$ immediately before contact line jumps occur. The drop base length 
and contact angle as a function of volume are plotted for the advancing contact line in 
Fig.~\ref{fig3}(b) and (c). The stick, slip, jump behaviour will be a recurrent theme for
all the substrates we consider.

\begin{figure} 
\begin{center}
\includegraphics[scale=0.8,angle=0]{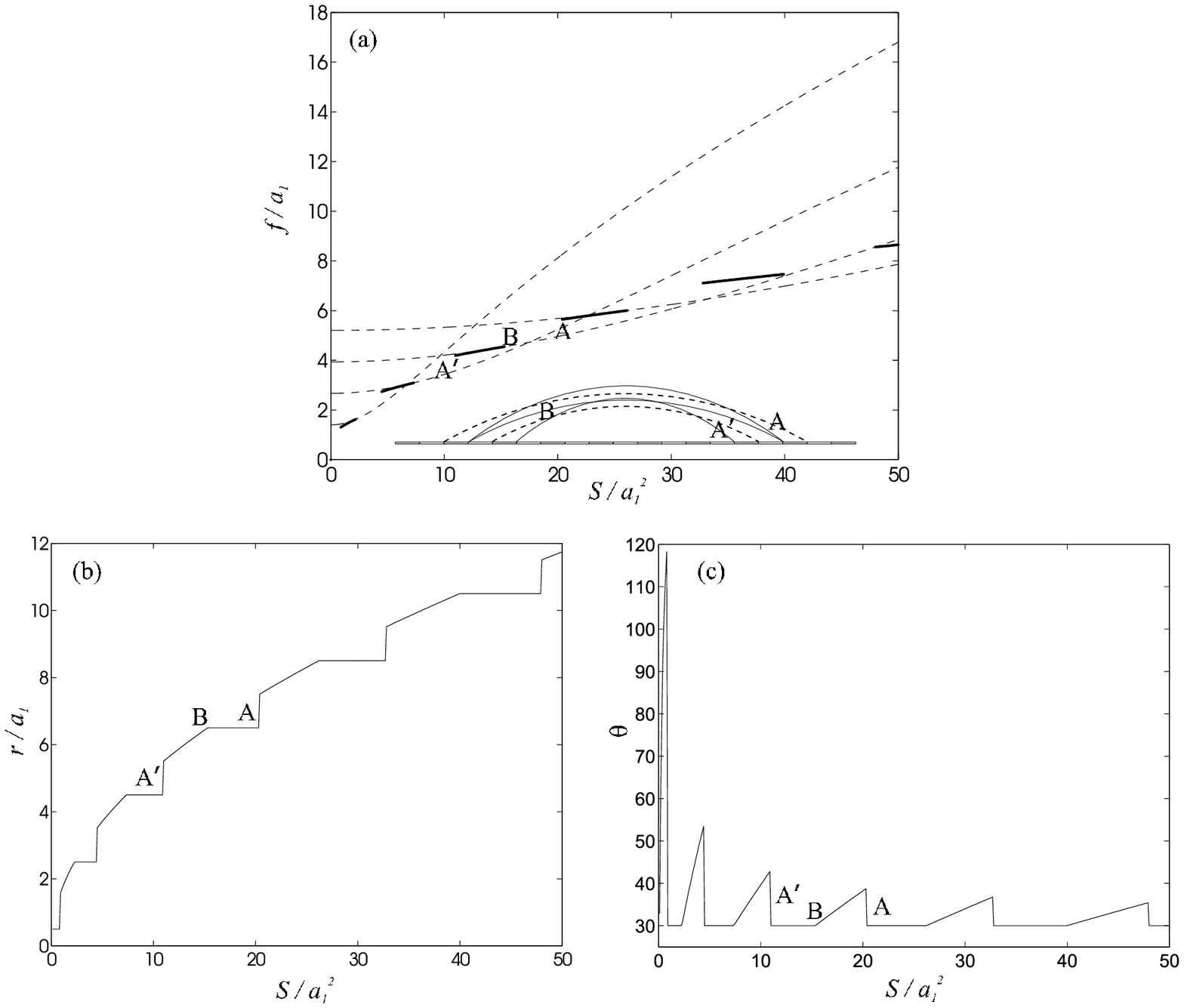}
\caption{
} \label{fig4}
\end{center}
\end{figure}

Similar reasoning can be applied to understand the dynamics of the
receding contact line. Free energy curves for this case are shown in
Fig. \ref{fig4}(a). (As this is the receding contact angle, the correct 
way to read the plot is from right to left.) Assume that the contact line 
is initially  placed on the hydrophobic stripe with $\theta=\theta_2$. 
As the volume of the drop is reduced, the contact line will recede smoothly to the
hydrophobic--hydrophilic border where it remains pinned until $\theta=\theta_1$ 
\cite{Iwamatsu1,Johnson1} (along AB in Fig. \ref{fig4}(a)).  The contact
line then continues to move along the hydrophilic stripe with contact
angle $\theta_1$ (BA$^\prime$). A sudden jump
occurs when it reaches the hydrophilic--hydrophobic border as the
drop can lower its free energy by dewetting the hydrophobic stripe (at
A$^\prime$ in Fig. \ref{fig4}(a)).  The inset in Fig. \ref{fig4}(a) show the profile
of the drop as its volume is reduced. The drop base length and contact angle as a 
function of volume are shown in Fig. \ref{fig4} (b) and (c). 

\begin{figure} 
\begin{center}
\includegraphics[scale=0.8,angle=0]{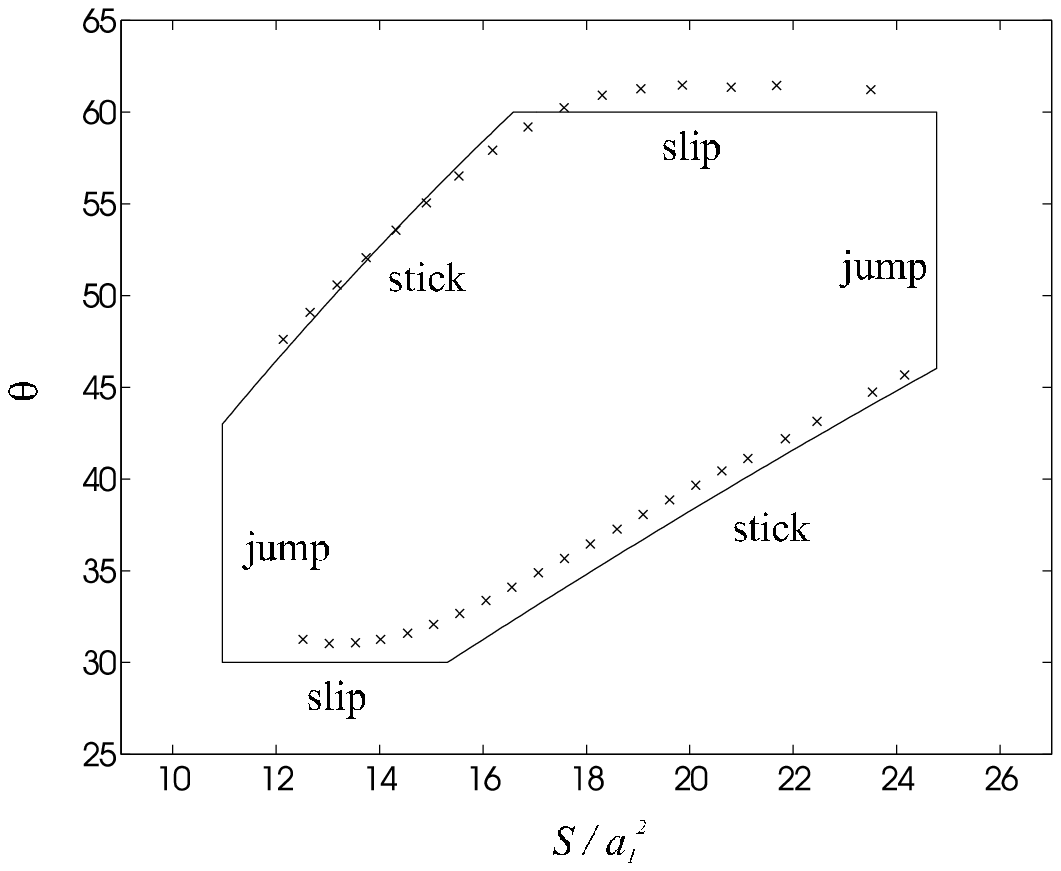}
\caption{
} \label{fig6}
\end{center}
\end{figure}

Fig. \ref{fig6} shows a typical hysteresis curve: the variation of the macroscopic
contact angle as the volume is increased, then decreased. Diagonal, horizontal,
and vertical lines correspond to the drop sticking, slipping, and jumping respectively.
The data points are results from the lattice Boltzmann simulations, while the solid line 
corresponds to the analytical calculations. The contact line dynamics is well captured 
by lattice Boltzmann simulations, though contact line jumps occur earlier in lattice 
Boltzmann simulations than one would expect from the analytical expressions. This is 
because the numerical interface has a finite width, typically $\sim 3$ lattice spacings.

We can conclude that for unforced hysteresis 
on a two dimensional chemically striped surface, the advancing contact angle 
$\theta_A = \theta_e|_{\mathrm{max}}$ and the receding contact angle
$\theta_R = \theta_e|_{\mathrm{min}}$. If we return to the free energy curves in
Fig~\ref{fig3}(a) and~\ref{fig4}(a), paths AB and BA$^{\prime}$ are reversible. 
Even though contact line pinning is observed there, the contact angle is unique 
for a given drop volume. It is the free energy jumps at A and A$^{\prime}$ that make 
contact angle hysteresis an irreversible phenomenon.


\section{Three dimensional drop on a chemically patterned surface}

In section III, we showed that $\theta_A = \theta_e|_{\mathrm{max}}$
and $\theta_R = \theta_e|_{\mathrm{min}}$ for a two dimensional drop
on a chemically striped surface. We now turn to three dimensions. Here
we find that the values of the advancing and receding  contact angles 
are strongly dependent on the details of the surface patterning. 
Therefore we shall focus on the
general aspects of the hysteresis, particularly those not captured by
the two dimensional model described in section III, rather than
quantitative details. The drops no longer form spherical caps so
analytic results are not possible without serious approximation. Our
arguments will be illustrated by numerical results obtained using lattice
Boltzmann simulations of the model described in Section II.

\begin{figure} 
\begin{center}
\includegraphics[scale=0.4,angle=0] {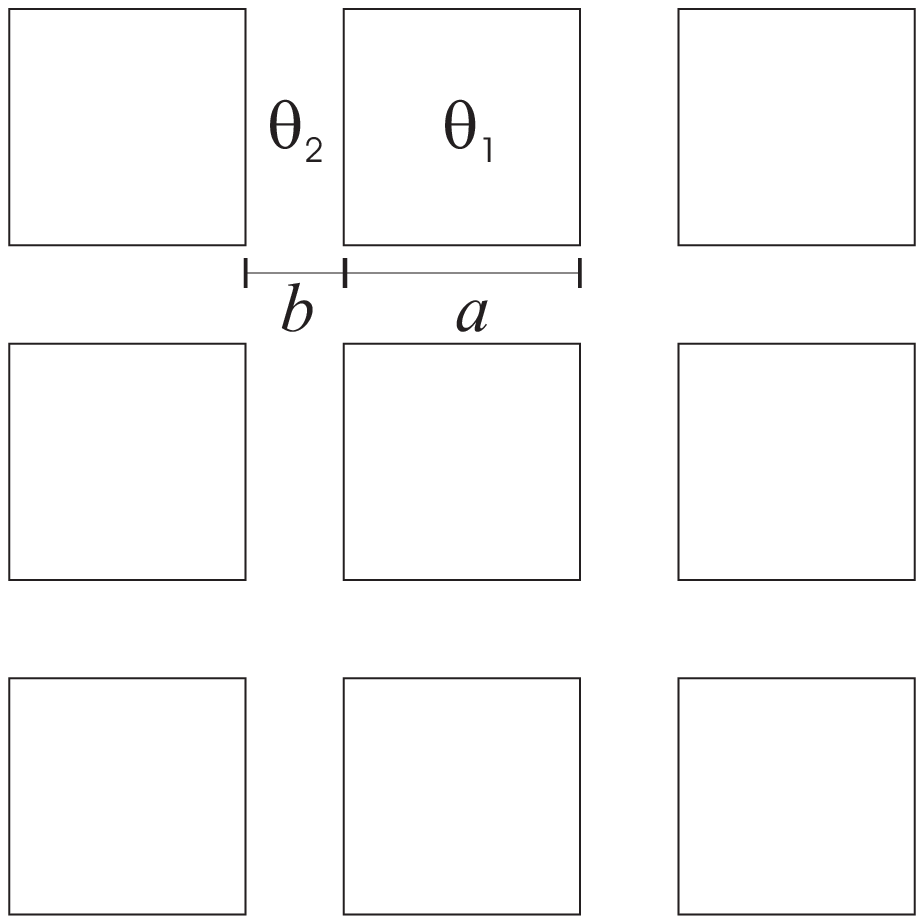}
\caption{
} \label{fig12}
\end{center}
\end{figure}

The chemically patterned surface we consider is shown diagrammatically
in Fig. \ref{fig12}. Squares of side $a=12$ are separated by a
distance $b=5$. In the first case that we consider (surface A), the
equilibrium contact angle of the squares is taken as
$\theta_1=110^{\mathrm{o}}$, while that of the channels between them is
$\theta_2=60^{\mathrm{o}}$. In the second case (surface B), we swapped
the values of the equilibrium  contact angles so that
$\theta_1=60^{\mathrm{o}}$ for the squares and
$\theta_2=110^{\mathrm{o}}$ for the channels. Due to the particular
choice of $a$  and $b$, the two surfaces have almost the same macroscopic 
contact angle, $\theta_{CB} \simeq 85.5^{\mathrm{o}}$ calculated using the
Cassie-Baxter formula
\begin{equation}
\cos{\theta_{CB}} = f_1 \cos{\theta_1} + f_2 \cos{\theta_2} \,  \label{CBC}
\end{equation}
which averages over the surface contact angles. $f_1$ and $f_2$ are the fractions
of the surface with intrinsic equilibrium contact angle $\theta_1$ and $\theta_2$
respectively. Therefore we might naively expect the two surfaces to have very similar behaviour. 
However, this is not the case unless the strength of the heterogeneities is below a
certain threshold \cite{deGennes2,deGennes1}, a condition which implies negligible contact angle 
hysteresis.

\begin{figure} 
\begin{center}
\includegraphics[scale=0.8,angle=0] {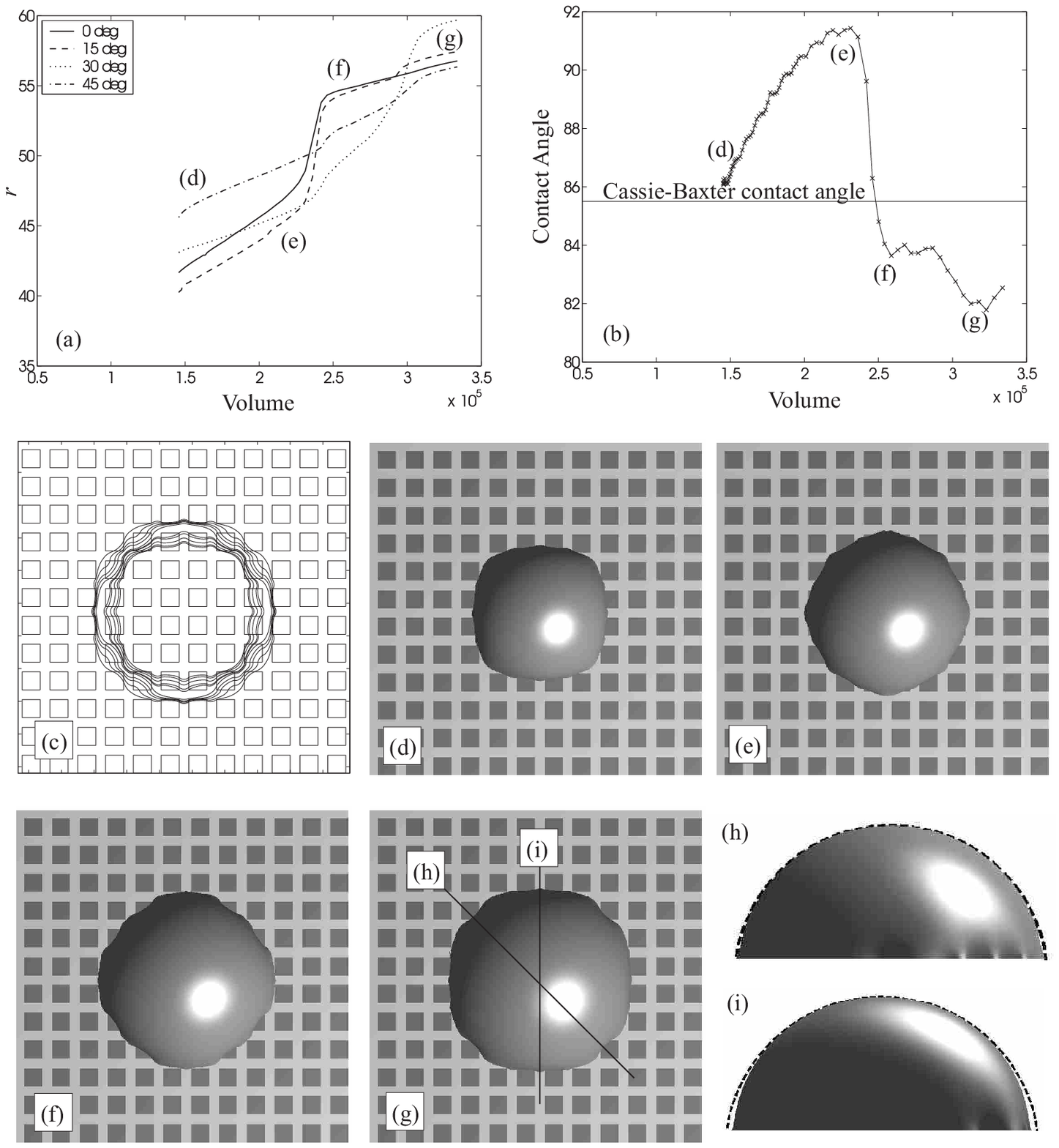}
\caption{
} \label{fig15}
\end{center}
\end{figure}

\begin{figure} 
\begin{center}
\includegraphics[scale=0.8,angle=0] {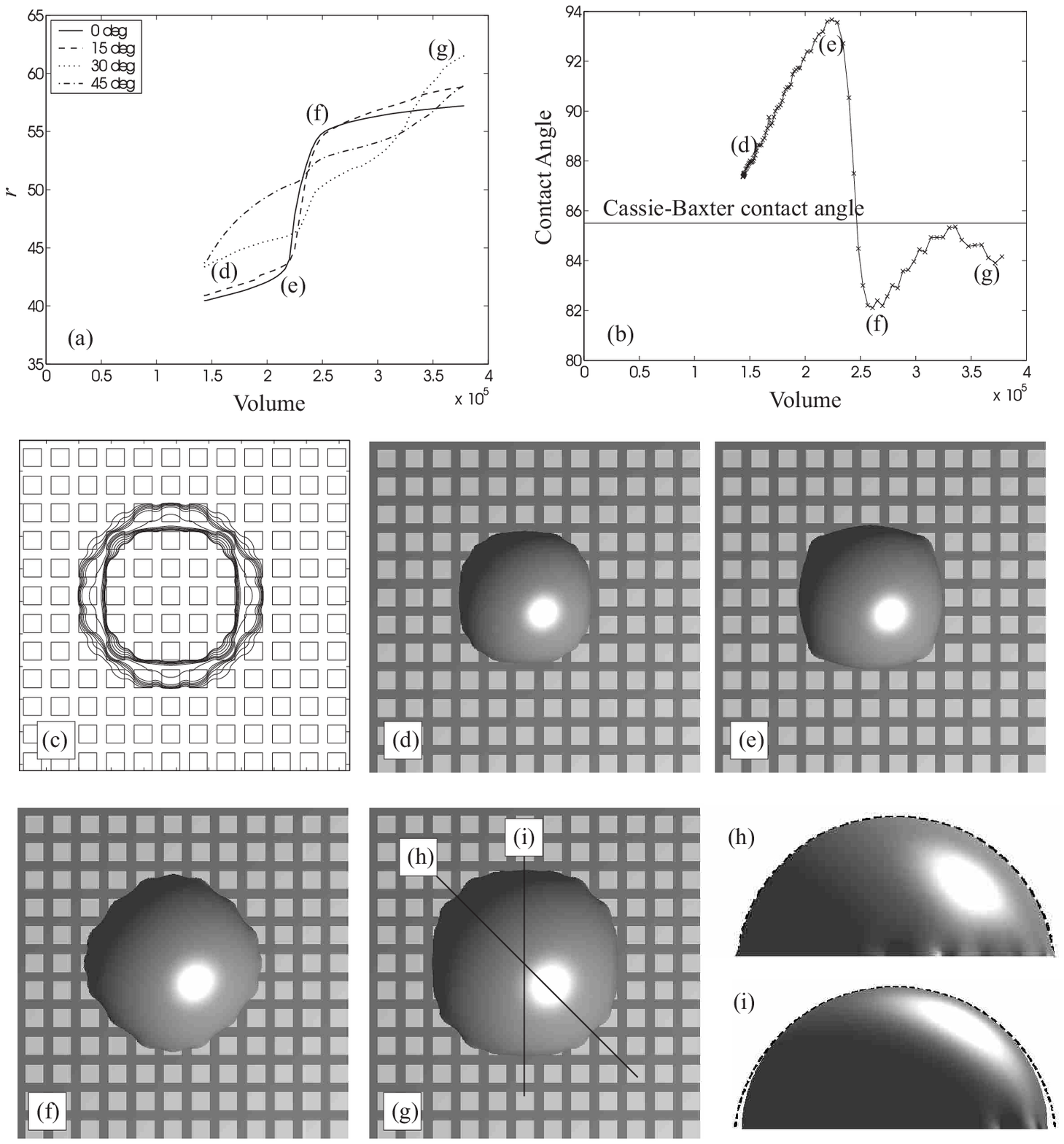}
\caption{
} \label{fig16}
\end{center}
\end{figure}

Consider first the advancing contact angle. From the
discussions in Section III, we know that contact line pinning 
occurs because there is an energy barrier for the drop to move 
onto a hydrophobic area, whilst the contact line jumps when 
it reaches a new hydrophilic region. Therefore we expect to 
observe stick behaviour when the contact line reaches unwetted 
hydrophobic parts of the surface, slip when it moves across partially covered 
hydrophobic portions, and jumps as the contact line reaches the hydrophilic areas.
This is indeed observed in the lattice Boltzmann simulations, as shown by 
the plots  of the base radius of the drop at different volumes in
Figs. \ref{fig15}(c) and \ref{fig16}(c) for surfaces A and B
respectively. 

Since the contact line is no longer circular the local contact angle varies
along the contact line and the notion of a macroscopic contact angle pertaining 
to the whole drop is not well defined. However, a popular approach (e.g. \cite{Chatain1})
is to match the drop profile far away from the substrate to a spherical cap to obtain
the drop radius of curvature, and hence define a macroscopic
contact angle 
\begin{equation}
\cos{\theta_{\mathrm{macro}}} = (1-h/R) \, , \label{eq18}
\end{equation}
where $h$ is the drop height and $R$ is the radius of curvature at the
top of the spherical cap. If the length scales of the heterogeneities are
of order the interface width then one expects Eq. (\ref{eq18}) to
give a value of $\theta_{\mathrm{macro}}$ equal to the Cassie-Baxter contact angle. 
However, as we see in Figs. \ref{fig15} -- \ref{fig16}, this is not necessarily
the case for the length scales considered here. Indeed, in some of
the metastable states the drop shape is far from a spherical cap 
and therefore it makes little sense to define a macroscopic contact
angle. As a guide to the accuracy of the definition (\ref{eq18}), 
Figs. \ref{fig15} -- \ref{fig16} (h--i) show the side views of the largest (for advancing contact angle)
or smallest (for receding contact angle) simulated drops at two different viewing angles, 
with the dashed lines showing the corresponding spherical fit. In general, the fit is better for the diagonal cross sections as compared to the
vertical or horizontal cross sections. This is because along the diagonal direction it is very disadvantageous for the contact line to lie
entirely in the hydrophilic region. As a result, contact line pinning is weaker, and the drop interface is more
able to take a circular shape. The top view of the drops are shown in 
Figs. \ref{fig15} -- \ref{fig16} (g). 

With these cautions the macroscopic contact angle, defined by Eq. \ref{eq18}
is plotted as a function of drop volume in Figs. \ref{fig15} -- \ref{fig16} (b). Fig.
\ref{fig15} gives the results for surface A and Fig. \ref{fig16} for surface B.
The overall behaviour of the curves is reminiscent of that for the two
dimensional case in Fig.~\ref{fig3}. However changes in slope are rounded and
there is secondary structure that depends on the details of the
surface patterning, and on the shape and position of the contact line. 
This is an important new feature of the three dimensional geometry; 
that the slip-stick-jump behaviour can occur at different volumes in 
different directions, leading to secondary structure in the base radius and
contact angle plots. This underlines the importance of using multiple viewing 
angles when measuring the advancing and receding contact angles on patterned 
surfaces. 

The behaviour of the different directions is, however, correlated. This can be seen in 
Figs. \ref{fig15}(a) and \ref{fig16}(a), where the variation of the interface position
is plotted against volume at several angles with respect to the lattice axes. 
For example, any variation in the drop radius at $0^{\mathrm{o}}$ is typically 
accompanied by a similar variation at $15^{\mathrm{o}}$, although the drop radius at $45^{\mathrm{o}}$
does not necessarily follow the same trend. Essentially this more complex motion of the 
drop in three dimensions occurs because the contact line lies on both hydrophobic and hydrophilic portions 
of the surface to avoid large surface distortions which would lead to a large liquid--gas 
surface free energy penalty. The shape of the contact line is thus very 
complicated and can be approximated analytically only in a few special cases \cite{Schwartz1,Schwartz2}. 
Nevertheless, as one would qualitatively expect, the simulation results show that the sections 
of contact line which lie on hydrophobic areas are concave whilst those on hydrophilic regions
are convex.

We now compare the behaviour on surfaces A and B. Results for surface A 
are shown in Fig. \ref{fig15}. As its volume increases the drop wishes 
to move onto the hydrophilic channels between the hydrophobic patches, 
but prefers to also cover part of the hydrophobic regions to avoid significant 
interface distortion. As a result, the drop tends to be facetted. The first, 
larger, jump in contact angle in Fig. \ref{fig15}(b) occurs when the 
contact line reaches the hydrophilic channels in the vertical and
horizontal directions. The contact line moves more easily in the 
diagonal directions and the second, smaller, jump is due to the 
sudden movement of the contact line along the diagonal. 

Contact line pinning is more pronounced on surface B (see Fig. \ref{fig16}). 
There is a relatively high free energy barrier to move onto the hydrophobic channels 
and hence contact line jumps occur at higher contact angles. As a corollary of this 
the free energy release and the jump in contact angle upon depinning is also higher 
for surface B. From Figs. \ref{fig15} and \ref{fig16}, the jumps can be read off as  
$8^{\mathrm{o}}$ and $11.5^{\mathrm{o}}$ for surfaces A and B respectively. Indeed 
even the order of the stick-slip-jump behaviour is not neccesarily the same. For the 
simulations we have presented in Figs. \ref{fig15} and \ref{fig16}, the order is 
stick-slip-jump on surface A, whereas on surface B, it is stick-jump-slip.

\begin{figure} 
\begin{center}
\includegraphics[scale=0.8,angle=0] {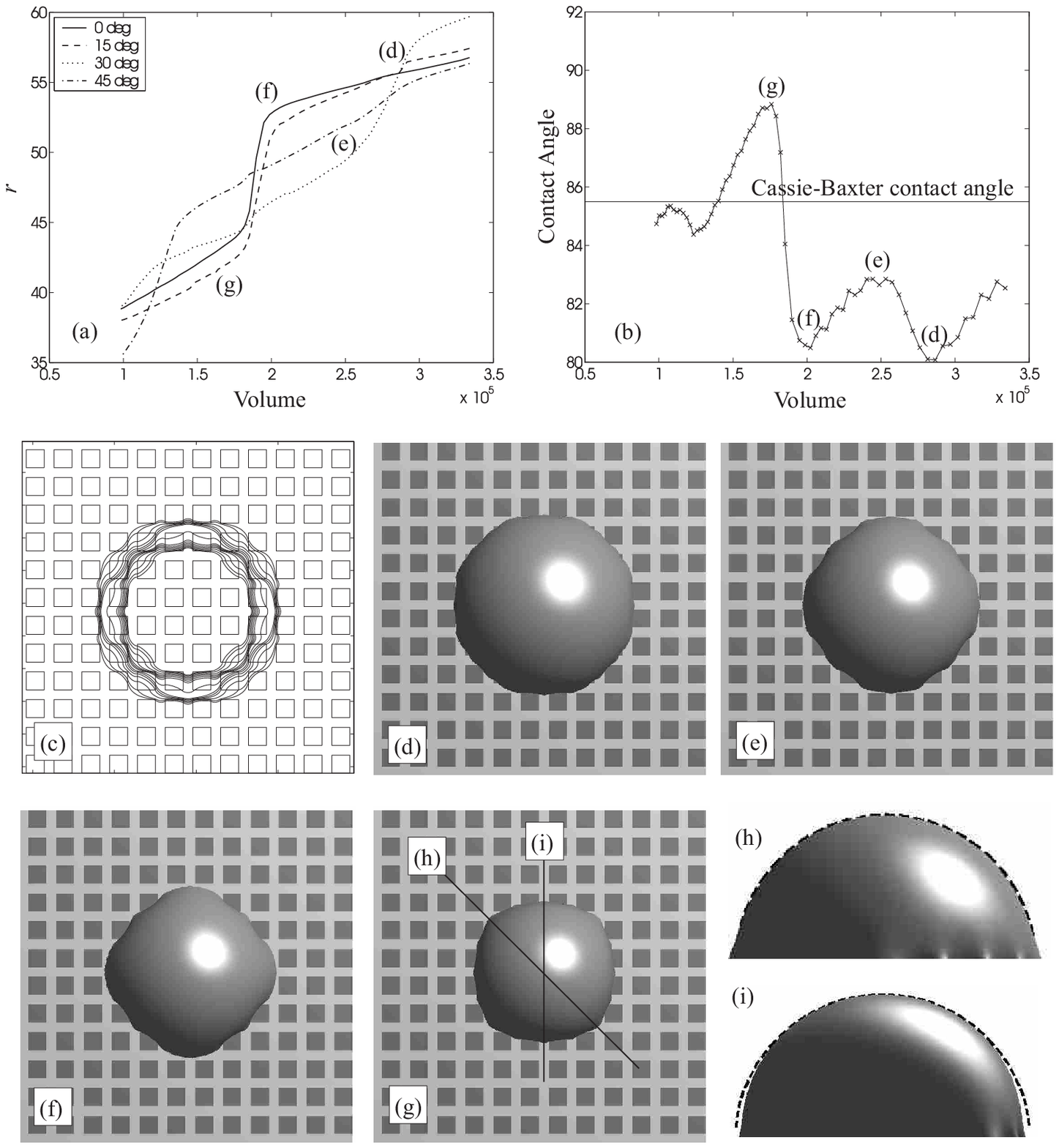}
\caption{
}  \label{fig22}
\end{center}
\end{figure}

\begin{figure} 
\begin{center}
\includegraphics[scale=0.8,angle=0] {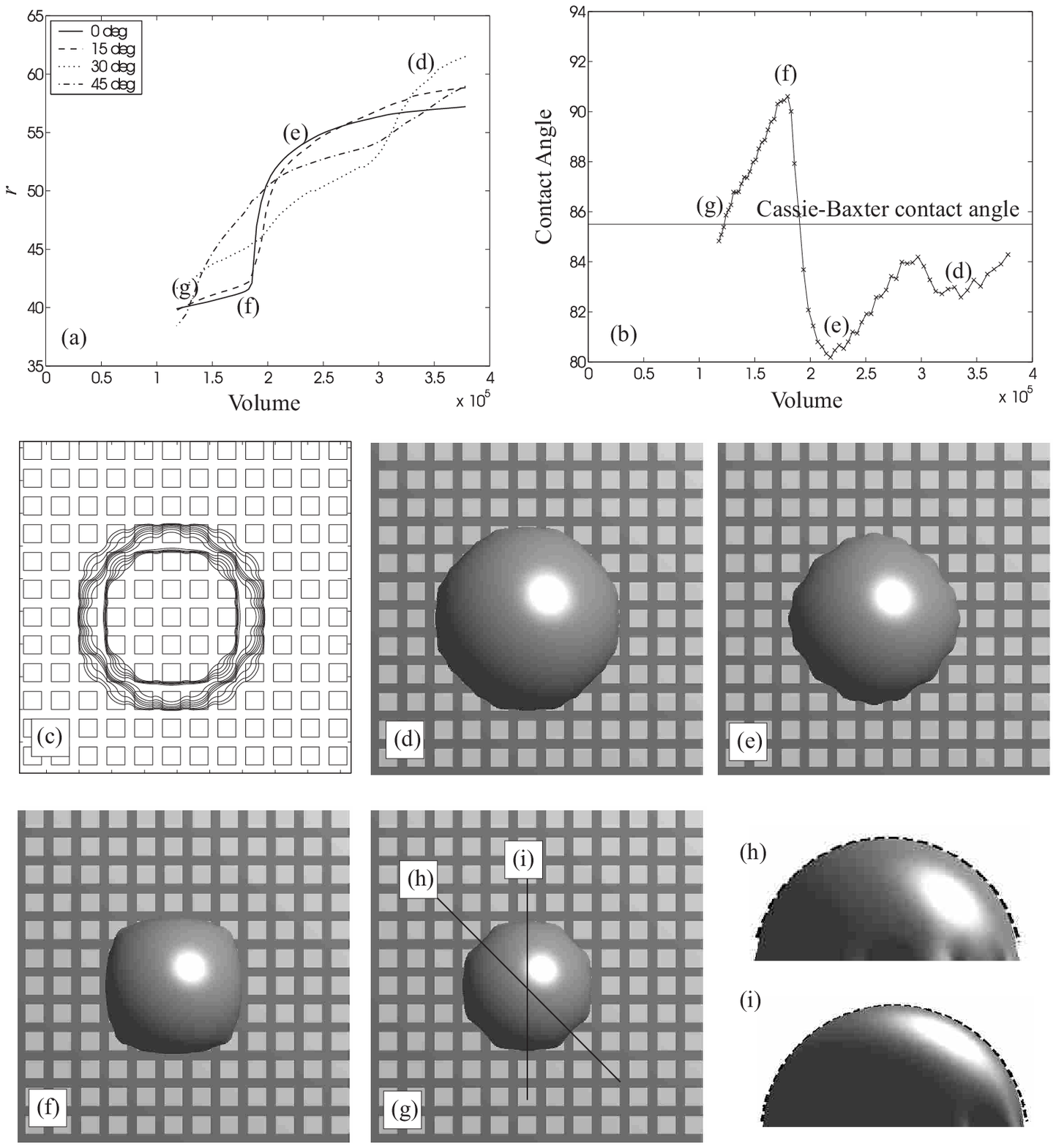}
\caption{
} \label{fig24}
\end{center}
\end{figure}

Similar results for the receding contact angles are presented in Figs. \ref{fig22} and \ref{fig24} for surfaces
A and B respectively. Comparing the contour plots for the advancing and receding contact line, we find that 
contact line pinning occurs at very similar positions: for surface A, when the drop tries to dewet the hydrophilic 
channel, and for surface B, as it dewets the square hydrophilic patches. Since the free energy barrier is highest in 
the vertical or horizontal directions, the primary features of the curves are caused by contact line pinning in 
these directions. The secondary features are due to the contact line dynamics in the diagonal directions. Again, 
this is similar to the behaviour as the drop advances. It is important to realise, however, that this is only
a qualitative similarity. The receding jumps occur at different volumes and have magnitudes  
$8.5^{\mathrm{o}}$ and $10.5^{\mathrm{o}}$ for surfaces A and B respectively. Figs. \ref{fig15}(b) and \ref{fig22}(b) 
also provide a clear example that the secondary structures can be different for the advancing and receding contact angles.

In two dimensions the advancing and receding contact angles were $\theta_e|_{\mathrm{max}}$ and $\theta_e|_{\mathrm{min}}$ 
(here $110^{\mathrm{o}}$ and $60^{\mathrm{o}}$). In three dimensions, these values are reduced by the additional surface
free energy penalty caused by distortions of the drop. On surfaces A and B, $\theta_A$ are $91.4^{\mathrm{o}}$ and 
$93.7^{\mathrm{o}}$ respectively, bounded by $\theta_e|_{\mathrm{max}}$ and the Cassie-Baxter contact angle $\theta_{CB}$.
The receding contact angles are $80.1^{\mathrm{o}}$ and $80.2^{\mathrm{o}}$ for surfaces A and B and are bounded by 
$\theta_e|_{\mathrm{min}}$ and $\theta_{CB}$. We therefore find that contact angle hysteresis is slightly larger for surface B 
($13.5^{\mathrm{o}}$) than for surface A ($11.3^{\mathrm{o}}$). We expect this discrepancy to become larger if the wettability 
contrast between the hydrophilic and hydrophobic patches is increased and future work will explore both this and the dependence 
of contact angle on patch size. The initial  positioning of the drop with respect to the hydrophobic grid will change the 
quantitative details of the behaviour, but we expect the qualitative descriptions we have here to remain true.


\section{Two dimensional suspended drop on a topologically patterned surface}

\begin{figure} 
\begin{center}
\includegraphics[scale=1.0,angle=0]{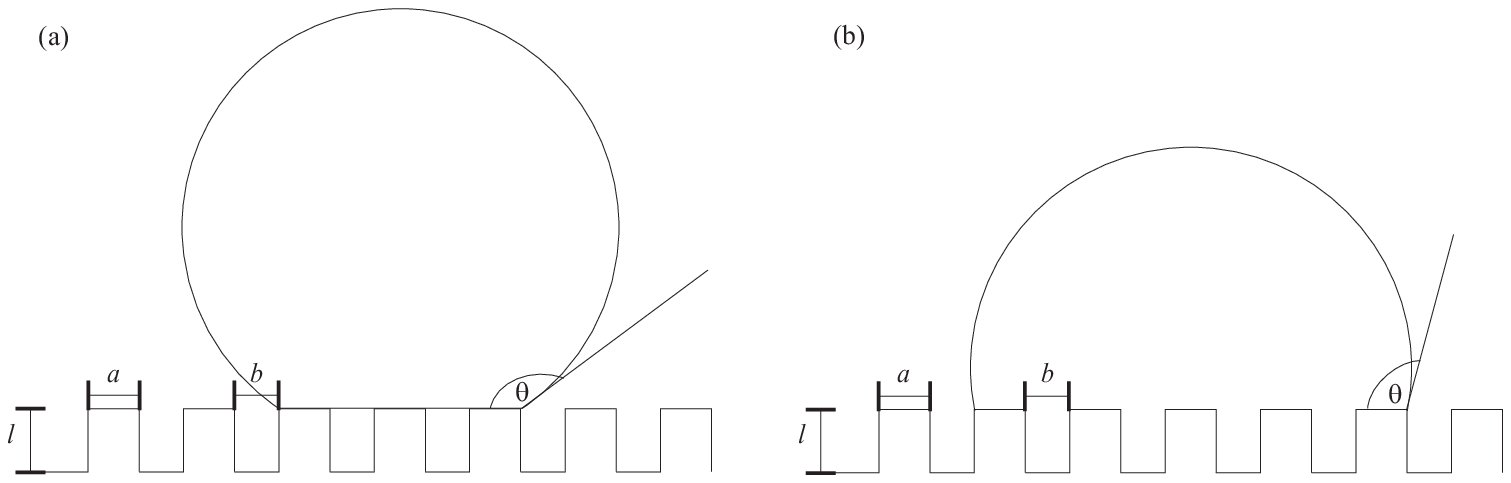}
\caption{
} \label{fig2}
\end{center}
\end{figure}

We now consider unforced hysteresis on a topologically patterned surface that consists of an array of posts as shown in
Fig. \ref{fig2}; this is a typical fabricated superhydrophobic surface. It is well-established that such topological 
heterogeneities amplify the hydrophobicity of the surface \cite{Quere1,Quere2}. There are two ways in which this can occur. When the drop 
is suspended on top of the surface roughness, as shown in Fig. \ref{fig2}(a), the substrate is effectively a composite of liquid-solid 
and liquid-gas sections. If the length scale of the posts is much smaller than the drop size, averaging over the corresponding
interfacial energies is equivalent to putting $\theta_1=\theta_e$ (of the posts) and $\theta_2=180^{\mathrm{o}}$ in the 
Cassie-Baxter equation (\ref{CBC}) to give a macroscopic contact angle 
equation \cite{Cassie}
\begin{equation}
\cos{\theta_{CB}} = f \cos{\theta_e} - (1-f) \, , \label{CB}
\end{equation}
where $f$ is the solid (area) fraction of the substrate. On the other hand, if the liquid drop fills the space between the
posts, as shown in Fig. \ref{fig2}(b), the liquid-solid contact area is increased by a roughness factor $r$, and the macroscopic contact 
angle is given by the Wenzel equation \cite{Wenzel}
\begin{equation}
\cos{\theta_{W}} = r \cos{\theta_e} \, . \label{W}
\end{equation}
This is called the collapsed state. Both the Wenzel and Cassie-Baxter formulae give reasonable estimates for the values of 
the contact angle. However, as they only take account of the average properties of the susbtrate and do not address the problem 
of multiple local free energy minima, their relevance in understanding contact angle hysteresis will be limited.

In this section, we concentrate on a two dimensional suspended drop, showing that the advancing and receding contact angles 
are $180^{\mathrm{o}}$ and $\theta_e$ respectively \cite{Huh2}. We then, in section VI, treat the two-dimensional collapsed drop. Here 
the hysteresis is larger with $\theta_A$ and $\theta_R$ equal to $180^{\mathrm{o}}$ and $\theta_e-90^{\mathrm{o}}$. The 
extension to three dimensions will be discussed in section VII and VIII for the suspended and collapsed drops respectively.

As the drop retains the shape of a circular cap above the posts, the free energy calculations described in
Section~\ref{chem2D} can be adapted to explore the way in which the contact line moves as the volume of the drop is slowly 
increased or decreased. The argument is analogous to that for the chemically striped surface if we take $\theta_1$ to be the 
equilibrium contact angle of the posts and $\theta_2$ to be the liquid--gas interface contact angle, which is equal to 
$180^{\mathrm{o}}$. Hence we expect the advancing and receding contact angles to be $\theta_A = \theta_e|_{\mathrm{max}} = 180^{\mathrm{o}}$
and $\theta_R = \theta_e|_{\mathrm{min}} = \theta_e|_{\mathrm{post}}$ respectively. For a finite value of base radius $r$, the contact
angle will reach $180^{\mathrm{o}}$ only as $R\rightarrow\infty$. As a result, the contact line will remain pinned indefinitely at the
outer edge of a given post as the drop volume is increased.

\begin{figure} 
\begin{center}
\includegraphics[scale=1.0,angle=0]{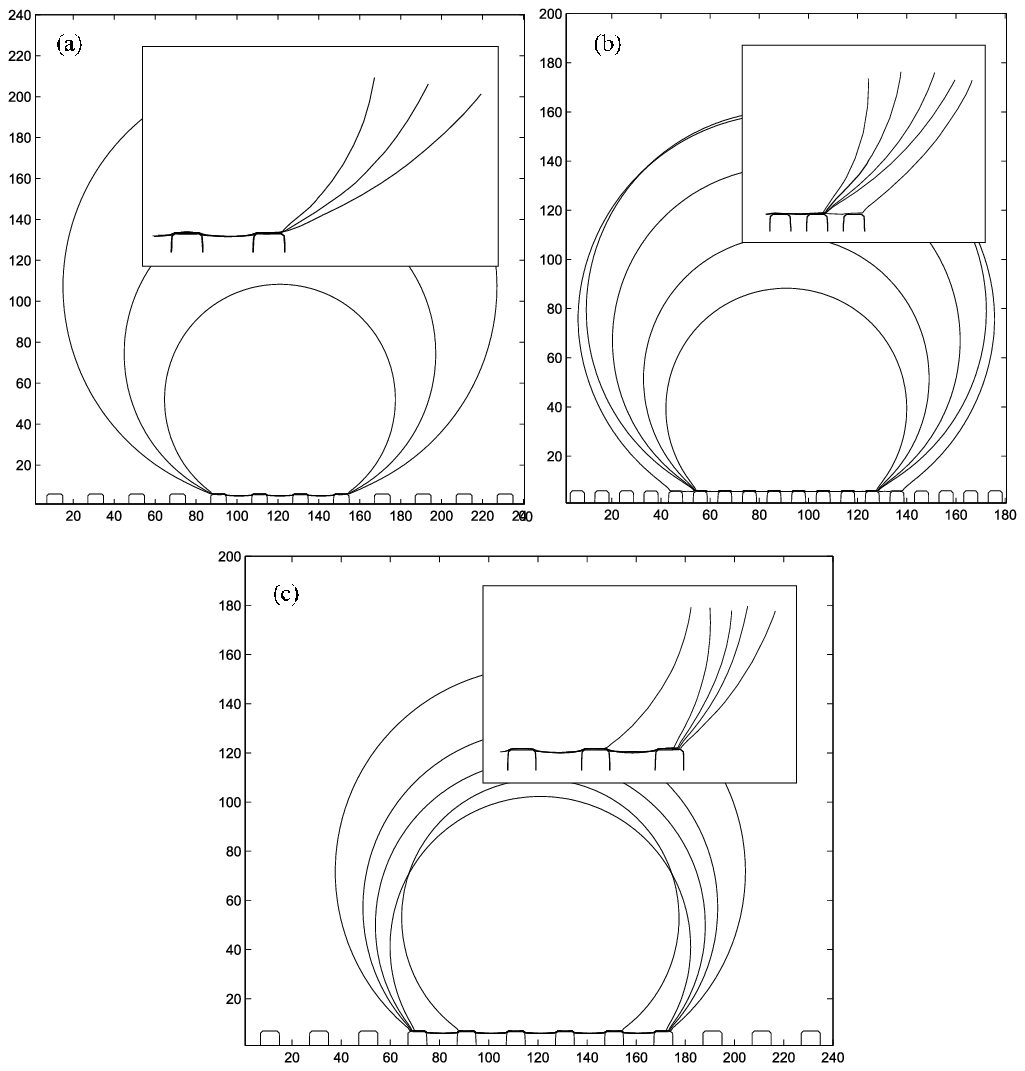}
\caption{
} \label{fig9}
\end{center}
\end{figure}

These predictions are consistent with lattice Boltzmann simulations as shown in Fig. \ref{fig9}(a).
In this set of simulations, we used a post width $a=7$, post
separation $b=13$, and an equilibrium contact angle $\theta_e = 120^{\mathrm{o}}$.
The drop volume increased by about a factor 4, and the contact angle to $162^{\mathrm{o}}$,
and no transition was observed. At this point it was no longer possible to
run the simulation as the drop filled the
simulation box. As the drop volume was decreased (Fig.~\ref{fig9}(c)) the
drop jumped between posts at $\theta \sim 120^{\mathrm{o}}$.
We do not find that the receding drop establishes its equilibrium
contact angle on the side of the asperities as proposed by Extrand
\cite{Extrand}.

If the separation between the posts was decreased to $b=5$ (with $a=5$), however,
we did observe a contact line jump from one post to another as shown in  
Fig. \ref{fig9}(b). The contact angle of the drop just before the
jump was $155^{\mathrm{o}}$. This is due to the diffuse liquid--gas
interface in the lattice Boltzmann simulations which allows the
interface to probe the existence of the neighbouring post and hence lower
its free energy by jumping across to it. 
To check this interpretation we ran a simulation increasing $\kappa$
from $0.004$ to $0.006$ corresponding to a wider interface. 
This led to a decrease in the advancing contact angle $\theta_A$ from $155^{\mathrm{o}}$
to $152^{\mathrm{o}}$. 

In the model, the interface width is comparable to the dimensions of the posts, 
a problem inherent in mesoscale simulations of multiphase fluids. This is not
true for typical superhydrophobic surfaces where the posts are $\sim \mu$m in 
size. Nevertheless, in real systems, there are likely to be mechanical vibrations 
\cite{Garoff1}, surface imperfections, and thermal fluctuations \cite{Kong}, which 
will cause the contact line to feel the neighbouring posts. Gravity will also 
lower the liquid--gas interface and hence makes it easier to touch the neighbouring posts.

\section{Two dimensional collapsed drop on a topologically patterned surface}

\begin{figure} 
\begin{center}
\includegraphics[scale=1.0,angle=0] {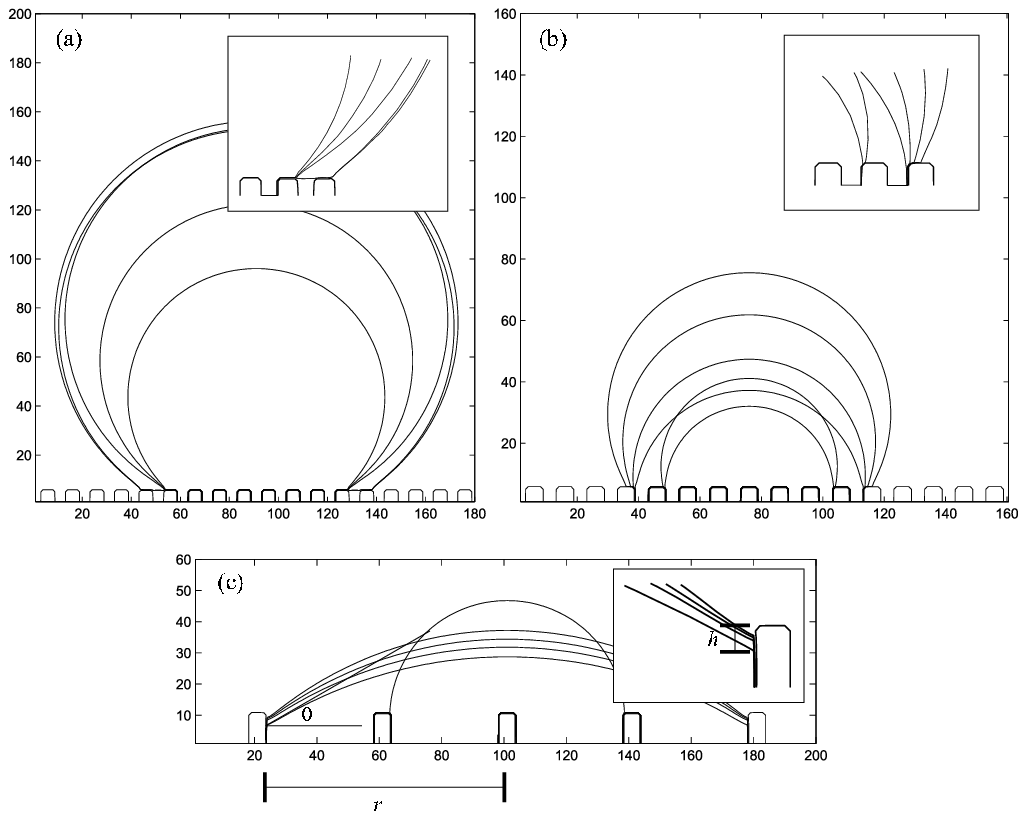}
\caption{
} \label{fig11}
\end{center}
\end{figure}

We now discuss unforced hysteresis for a cylindrical collapsed drop where the gaps
between the posts are filled with liquid. When the drop volume is increased
the advancing contact angle is $180^{\mathrm{o}}$ and the drop behaves in the same way as for the
suspended state. This is because locally, in the vicinity of the
contact line, the drop has no information as to whether it is in the
collapsed or suspended state. Indeed, in the two dimensional simulations we 
have run, if the contact line does advance, the grooves are not filled by the liquid drop as the contact line advances
from one post to the next, in contradiction to the work of Li {\em{et al.}} \cite{Amirfazli}, 
which assumes that the gap is filled as the contact line advances. The results are 
shown in Fig. \ref{fig11}(a) for $a=5$, $b=5$, $l=5$, and $\theta_e = 120^{\mathrm{o}}$,
the same parameters as in Fig. \ref{fig9}(b). In a perfect system there should be no 
advancing jump between posts. We see a transition at $\theta = 154^{\mathrm{o}}$ for the
collapsed drop because of the diffuseness of the interface. This is consistent with the results in 
Fig. \ref{fig9}(b) where a contact line jump was seen at $\theta = 155^{\mathrm{o}}$
for the suspended drop. 

Next we consider the way in which the contact line recedes as the drop
volume decreases. The typical behaviour is shown
in Fig. \ref{fig11}(b). As for the suspended drop, the contact
line is pinned at the outer edge of a post until $\theta=\theta_e$. It
then retreats smoothly across the post. However, unlike the suspended
case, the contact line is pinned again at the inner edge of the
posts. This happens because the free energy gain from the
reduced liquid--gas interface of the circular cap is not large enough
to dewet the side walls of the posts. 

To demonstrate this we recall that drop free energy has two important 
contributions. The first comes from the liquid--gas interface of the
circular cap, while the second is from the liquid--solid boundary.
\begin{equation}
f \equiv F/\gamma_{LG} = \frac{2\,r\,\theta}{\sin{\theta}} + 2\,h\,\cos{\theta_e} \, ,
\end{equation}
where $r$, $\theta$, and $h$ are respectively the base radius, the macroscopic contact angle, and the distance
from the inner edge of the posts to the contact line. These are illustrated in Fig. \ref{fig11}(c). Taking the
first derivative of $f$ gives
\begin{equation}
df = \left[ \frac{2\,r}{\sin{\theta}} - \frac{2\,r\,\theta\,\cos{\theta}}{\sin^2{\theta}} \right] d\theta + 2\,\cos{\theta_e}\,dh \label{eq19} \, .
\end{equation}
For a drop of a constant volume 
\begin{equation}
S = r^2 \, \frac{\theta-\sin{\theta}\cos{\theta}}{\sin^2{\theta}} - 2 \, r \, h \,  \nonumber
\end{equation}
and hence $dh$ to $d\theta$ are related by
\begin{equation}
dh = - r \, \frac{\theta\cos{\theta}-\sin{\theta}}{\sin^3{\theta}} \, d\theta \, . \label{dhdt}
\end{equation}
Subsituting Eq. (\ref{dhdt}) into Eq. (\ref{eq19}) gives
\begin{equation}
df = 2 \, (\sin{\theta}+\cos{\theta_e}) dh \, .
\end{equation}
The condition $df/dh > 0$ implies that pinning will hold at
the top of the posts as long as $\theta > \theta_e - 90^{\mathrm{o}}$. 
This is best seen in Fig. \ref{fig11}(c) where the posts' separation
has been increased to $b = 35$ to facilitate the measurement of the receding
contact angle. The drop is found to recede at $32^{\mathrm{o}}$, consistent 
with the expected analytical result $\theta_e-90^{\mathrm{o}}=30^{\mathrm{o}}$.
We note that this is equivalent to the criteria first 
proposed by Gibbs \cite{Gibbs} and confirmed experimentally by Oliver 
{\it et\ al.\ } \cite{Huh2}.

Once this condition is satisfied, further reduction in the drop volume 
will cause the drop to steadily dewet the side of the post. If the posts 
are short, the contact line will touch the base of the interstices and 
it will quickly retract to the neighbouring posts. If the posts are tall, 
however, the liquid--gas interface can intersect the corners of the 
neighbouring post, in which case satellite drops will be formed in between 
the posts. In either case, the contact line will typically reach the next
post and move to its inner edge without regaining a contact angle $\theta_e$. 
Therefore the behaviour will typically be stick-jump rather than stick-jump-slip. 
Furthermore, comparing this motion to that described for the suspended drop in
section V shows that, in this simple two dimensional model, the contact angle 
hysteresis is $90^{\mathrm{o}}$ larger for the collapsed state than for the 
suspended state.


\section{Three dimensional suspended drop on a topologically patterned surface}

The two dimensional model discussed in Section V and VI captured the phenomenon of contact line 
pinning on topologically patterned surfaces but the three phase contact line was only represented by two points at
the ends of the liquid--gas interface. Hence we were unable to
address the effect of distortions of the contact line caused by the
surface roughness. Therfore we now present results from three dimensional lattice
Boltzmann simulations aimed at investigating contact angle hysteresis
on three dimensional topologically patterned surfaces.

For the suspended state we used $a = 3$, $b = 7$, $l=5$, 
and $\theta_e=110^{\mathrm{o}}$ which, using the Cassie-Baxter
formula (\ref{CB}), gives an estimate of $\theta = 160^{\mathrm{o}}$ for the macroscopic 
contact angle. The very high value of the drop contact angle is challenging for 
the simulations because a small change in the contact angle requires a large 
change in the drop volume. Moreover, since one needs the drop to be suspended 
on a reasonable number of posts, a huge simulation box is needed. We used 168 x 168 x 168 lattice points
and the largest drop simulated had a volume of $16.8 \, \mathrm{x} \, 10^{5}$.

Small deviations in the drop shape near the surface (from a spherical cap) can cause 
large uncertainties in the contact angle measurements. The definition in Eq. (\ref{eq18}) 
gives $\theta = 180^{\mathrm{o}}$ for the drops modelled in this section. We therefore use 
\begin{equation}
\theta_{\mathrm{macro}} = 2 \tan^{-1}\left(\frac{H}{r_{\mathrm{max}}}\right)  \label{eq30}
\end{equation}
where $H$ is the height of the drop and $r_{\mathrm{max}}$ is the maximum base radius, 
as an alternative definition of macroscopic contact angle.

\begin{figure} 
\begin{center}
\includegraphics[scale=0.8,angle=0] {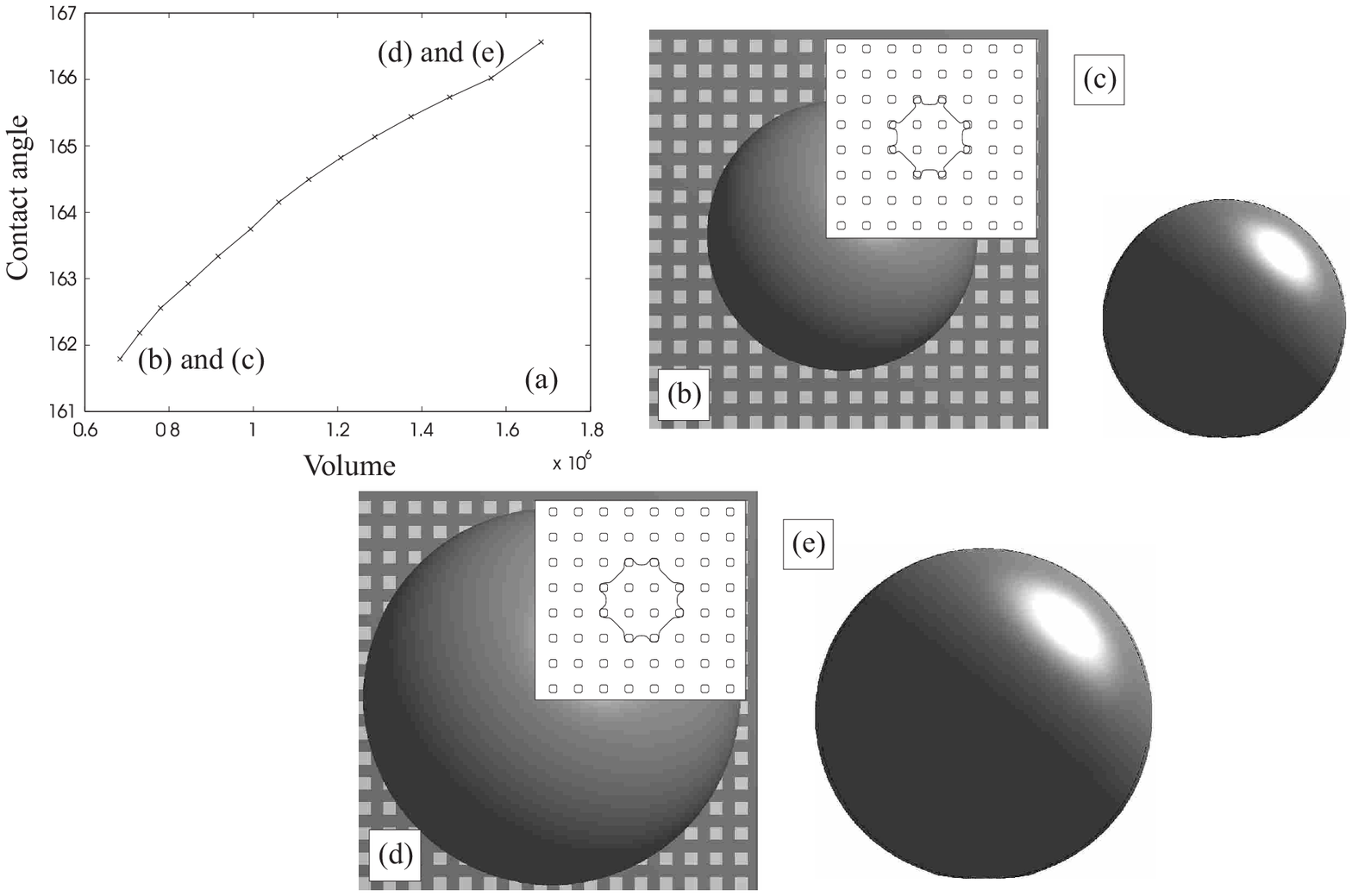}
\caption{
} \label{fig17}
\end{center}
\end{figure}

Fig. \ref{fig17} (a) shows the drop contact angle as a function of volume. In the simulation,
we have quasi-statically increased the volume from $6.8 \, \mathrm{x} \, 10^{5}$ to 
$16.8 \, \mathrm{x} \, 10^{5}$. This gives a macroscopic contact angle $\theta$ (Eq. (\ref{eq30}))
which ranges from $161.8^{\mathrm{o}}$ to $166.6^{\mathrm{o}}$. At least for the range of 
volumes simulated the contact line remains pinned as suggested by other groups 
\cite{McCarthy1,McCarthy2,McCarthy3,Chatain1,Dorrer} and consistent with our two 
dimensional arguments that the contact line jump occurs when 
$\theta\rightarrow 180^{\mathrm{o}}$ or when the interface is able 
to touch the neighbouring posts. In Fig. \ref{fig17}(b--e) the contour plot, top view, and side 
view of the drop at the beginning and at the end of the simulation are shown. 

\begin{figure} 
\begin{center}
\includegraphics[scale=0.8,angle=0] {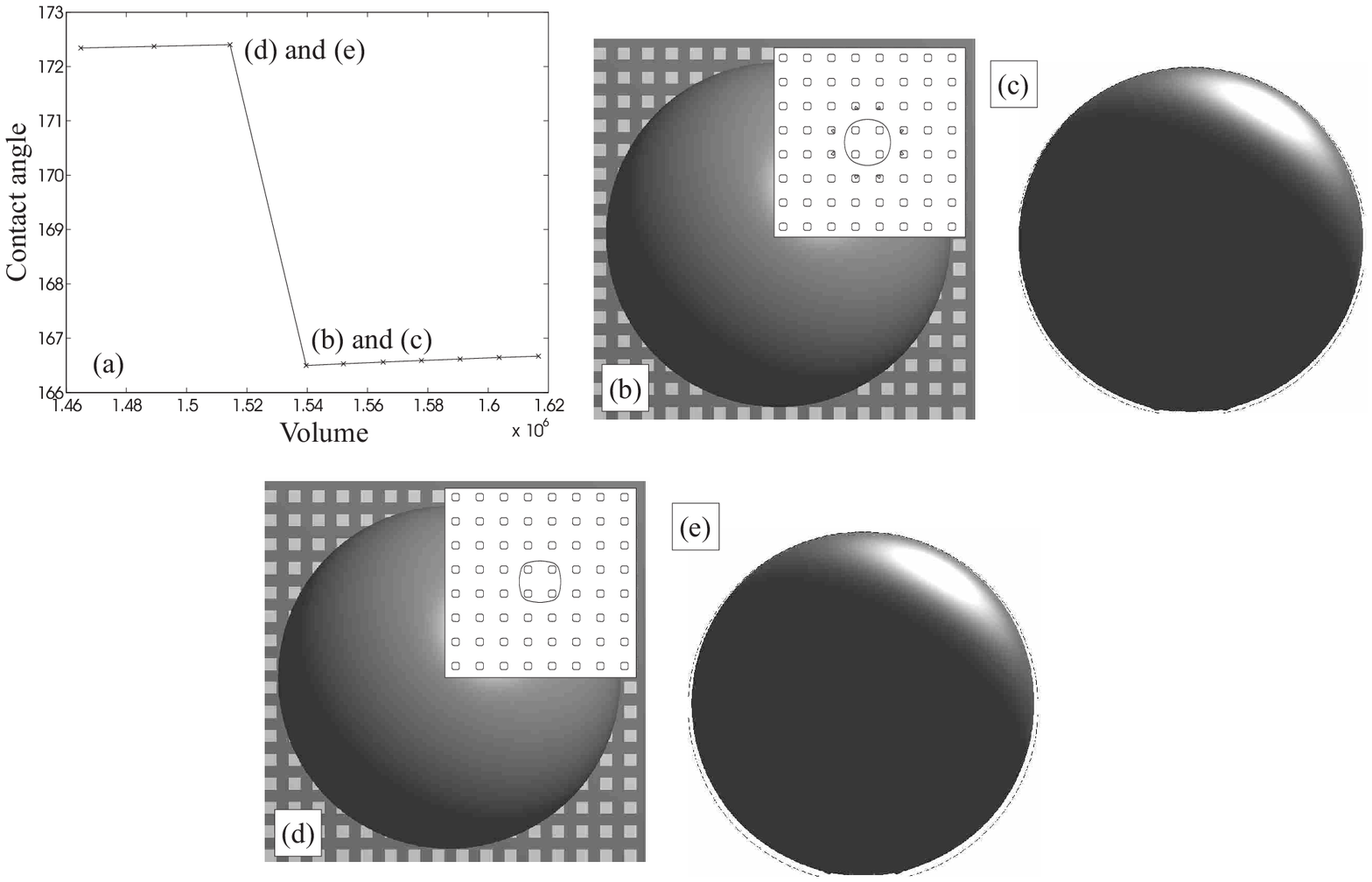}
\caption{
} \label{fig18}
\end{center}
\end{figure}

While the contact line is very difficult to advance, it is a lot easier to recede. 
We reduce the drop volume by reducing the liquid density by around $0.1\%$ and then 
let the system to relax to a (local) minimum energy state. For the parameters used here,
the receding contact line jump occurs at $166.5^{\mathrm{o}}$ ($V = 15.4 \, \mathrm{x} \, 10^{5}$), 
as shown in Fig. \ref{fig18}(a). This high value of the receding angle implies that the receding contact line is very 
unstable against small perturbations, highlighting the weak adhesion of suspended drops 
on superhydrophobic surfaces. Indeed even the small volume changes we used in the simulations
tended to make the drop detach slightly from the surface. The receding contact line can be made even less stable by 
reducing the solid fraction or by increasing the hydrophobicity of the posts \cite{McCarthy1,McCarthy2}. The contour plot, 
top view, and side view of the drop before and after the contact line jump are shown in
Fig. \ref{fig18}(b--e).


\section{Three dimensional collapsed drop on a topologically patterned surface}

We now turn our attention to the collapsed state where the space between the posts is 
filled with liquid. We use $a=4$, $b=6$, $l=5$, and $\theta_e = 120^{\mathrm{o}}$, 
which according to the Wenzel formula (\ref{W}) gives a contact angle $\theta_W=154^{\mathrm{o}}$.

\begin{figure} 
\begin{center}
\includegraphics[scale=0.7,angle=0] {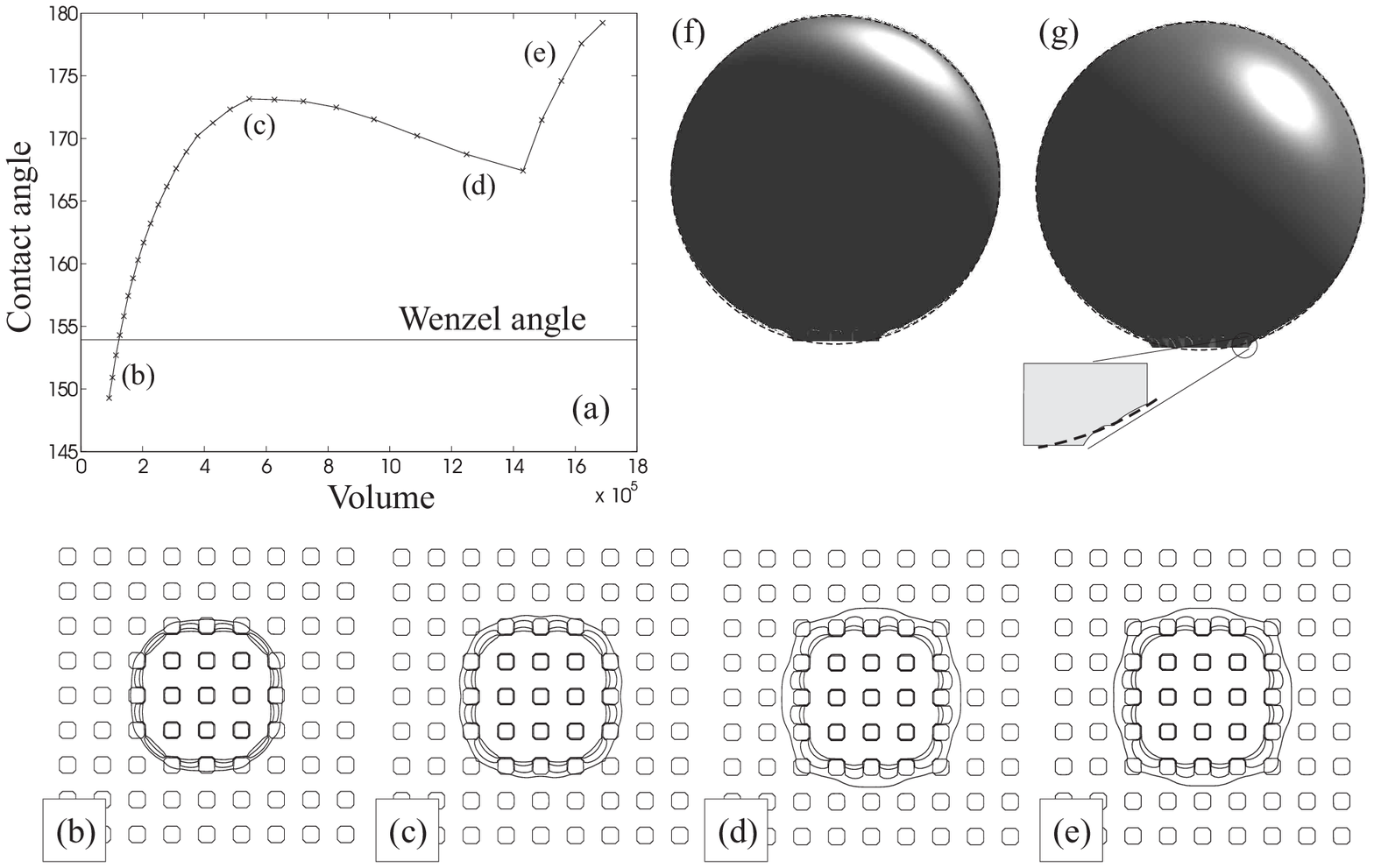}
\caption{
} \label{fig19}
\end{center}
\end{figure}

Fig. \ref{fig19}(a) shows the macroscopic contact angle of the drop, calculated using the definition
(\ref{eq18}), as a function of increasing volume. Initially the contact angle increases 
because, even though the contact line moves outwards in the diagonal direction (with respect 
to the posts), it remains pinned in the horizontal and vertical directions. This can be seen 
in the contour plots shown in Fig. \ref{fig19}(b--c). Once the drop has touched the four 
neighbouring posts along the diagonals, it wets the top of these posts (Fig. \ref{fig19}(d--e)) 
and this allows the contact angle to decrease slightly, by about $5^{\mathrm{o}}$. At this stage 
the contact line will again be pinned until it is energetically favourable to jump and wet the 
neighbouring posts in the vertical and horizontal directions. This will become possible when 
the drop can feel the neighbouring posts. We note that this contact line jump has not occured 
in the simulations we have presented here. Increasing the size of the simulation box was 
prohibitively expensive in computer time

Typical side views of the drops are shown in Fig. \ref{fig19}(f--g), with the dashed line showing 
the corresponding spherical fits. The fitted curves match well with drop profiles above the posts, but
not with the profiles in between the posts. This again emphasises that any definition of 
macroscopic contact angle is arbitrary and should be treated with caution. In this section, the 
contact angle is calculated at the intersection between the fitted curve and the base surface. 
If instead we take the intersection between the fitted curve and the top of the posts, the values 
of the contact angles would be lower. This will however, not change the trends we describe in this section.

In section VI, we found that, in two dimensions, the receding contact angle for the
collapsed state is $\theta_e - 90^{\mathrm{o}}$ because of the adhesion 
provided by the sides of the posts. In three dimensions, we expect the adhesion effect to remain
relevant. However, the corresponding energy barrier will now be lowered by the free energy costs 
due to the liquid--gas interfaces between the posts. 

\begin{figure} 
\begin{center}
\includegraphics[scale=0.7,angle=0] {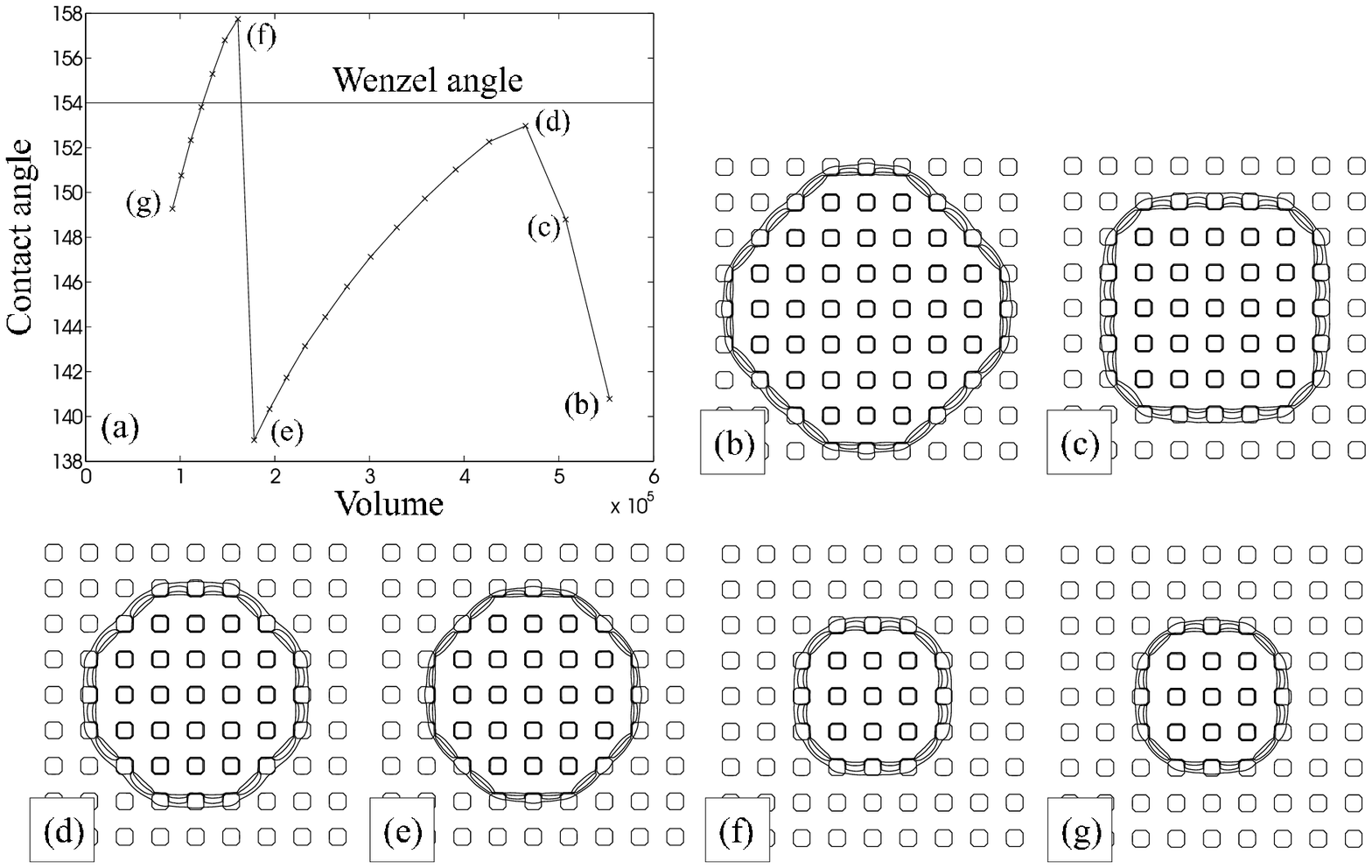}
\caption{
} \label{fig20}
\end{center}
\end{figure}

The drop's macroscopic contact angle as a function of decreasing volume is shown in Fig. \ref{fig20}(a). 
The general features of the curve are very similar to the other examples of receding contact line 
dynamics that we have discussed. The contact angle decreases slowly as volume is reduced 
(Fig. \ref{fig20}(d--e) and Figs. \ref{fig20}(f--g)) until any further reduction in drop volume 
makes it more favourable for the drop to depin its contact line (Figs. \ref{fig20}(b--d) and \ref{fig20}(e--f)). 
Naturally, as a result of the contact line jump, the contact angle increases very sharply. 
Taking the lowest (highest) value of the contact angle in Fig. \ref{fig20}(a) (Fig. \ref{fig19}(a))
as the receding (advancing) contact angle, we obtain $\theta_R=139^{\mathrm{o}}$, 
$\theta_A=179^{\mathrm{o}}$ and $\Delta\theta \sim 40^{\mathrm{o}}$. 


\section{Discussion}

In this paper we described contact angle hysteresis on both chemically patterned and 
superhydrophobic surfaces as the drop volume was quasi-statically increased or decreased. 
We first considered the two dimensional case of a cylindical drop. This is a useful 
baseline calculation as the models are tractable analytically and contain many features 
that carry over to three dimensions. It was also useful to check the results of lattice 
Boltzmann simulations against the two dimensional results. The simulation approach was 
then used to explore the full, three dimensional problem.

For chemically patterned surfaces in two dimensions the advancing (receding) contact 
angle is equal to the maximum (minimum) value of the equilibrium contact angle. In three 
dimensions these parameters act as bounds on the hysteresis. Extra free energy 
contributions, which result from the distortion of the surface, decrease the advancing, 
and increase the receding, contact angles, thus decreasing the contact angle hysteresis. 

For superhydrophobic surfaces in two dimensions the advancing contact angle is close to 
$180^{\mathrm{o}}$ for both suspended and collapsed drops. The drop will only advance when it is 
able to feel the neighbouring posts because of, for example, finite interface width or 
mechanical vibration. This remains true in three dimensions.

The receding contact angle is, however, very different for the suspended and collapsed 
states. In two dimensions, for suspended drops, it is equal to the contact angle of the 
posts $\theta_e$. For the collapsed state it is equal to $\theta_e - 90^{\mathrm{o}}$ because the 
interface needs to move down the sides of the posts to recede. In three dimensions these angles 
again provide limiting values. 

For a suspended drop, to obtain a low contact angle hysteresis, the receding angle needs to 
be driven as close to $180^{\mathrm{o}}$ as possible \cite{McCarthy1,McCarthy2}. This can be done by making $\theta_e$ as large 
as is feasible and then choosing the surface patterning to distort the drop so that the free 
energy barriers are lowered. For the collapsed state, contact line pinning is stronger and
hence we expect the receding contact angle to be lower and the contact angle hysteresis
to be larger. This is observed experimentally (e.g. \cite{Quere1,Quere2,McCarthy1})
and is related to the very different adhesive properties of the suspended and collapsed 
states. To increase the receding contact angle, one can increase $\theta_e$, 
increase the posts' separation, or decrease the posts' size. 

It is important to note that, because of the distorted drop shape, the advancing and 
receding contact angles depend on the direction in which the measurement is taken; indeed 
the concept of macroscopic contact angle is a matter of definition. They also depend 
sensitively on the details of the surface patterning and any average over the area of the 
patterning or over the angle along the contact line is not sufficient to predict the 
hysteretic properties of a surface. We also note that when the contact line jump 
involves more than one post (or chemical patch), it does not wet/dewet the posts simulatenously.
For the advancing drop, movement starts at the centre post, while for the receding drop, the
contact line first dewets the outer posts. This is in agreement with
\cite{deGennes2,Schwartz1,Schwartz2,Dorrer}.

It will be of interest to compare the results we have presented here for unforced
hysteresis, where the drop volume is changed quasi-statically, to what we will term
forced hysteresis, where the drop is pushed by, for example, a body force. The aim
is to slowly increase the force and discuss the value of the advancing and 
receding contact angles at which the drop first starts to move. Because we are 
working in the limit of zero velocity free energy arguments can still be used. 

On a flat, perfectly smooth surface the limiting force is zero and the drop will 
start to move as soon as it is pushed. On a patterned surface there will be pinning. 
The problem is, however, now much more complicated than for unforced
hysteresis as the free energy barriers, of which there might be many, depend 
not only on the local surface patterning, but also on the overall shape of the drop. This in
turn will depend on the size of the drop and the magnitude and details of the applied force. 
Therefore we expect that, in general, the advancing and receeding contact angles will differ 
from those of unforced hysteresis and that they will depend sensitively on the details of
the problem. Preliminary results indicate that the stick-slip-jump behaviour is still observed, 
both in two and three dimensions. In general the depinning of the front and back contact line 
occur at different times, as opposed to the simultaneous depinning seen in the unforced case. 

The aim will then be to consider dynamic hysteresis, the effect of free energy 
barriers on the motion of a drop. This is now a dynamical problem which will depend on 
the equations of motion of the drop as well as its static properties. For the results 
presented in this paper the full lattice Boltzmann formalism is not needed, except to follow the 
motion of a drop during a jump and to aid relaxation to equilibrium. However, we wanted 
to use it to allow an immediate generalisation to dynamical problems. A fuller investigation of 
forced and dynamic hysteresis will be presented elsewhere.

There are many other directions for future work. Among these are the effects of random 
patterning on the advancing and receding contact angle and any differences that may arise 
from having smooth, rather than abrupt, junctions between regions of different contact 
angle. It would also be interesting to perform simulations to more fully explore the dependence
of contact angle hysteresis on the size and spacing of the posts and chemical patches.


\section*{Acknowledgements}
We thank G. A. D. Briggs, A. Dupuis, G. McHale, D. Qu\'{e}r\'{e} and R. J. Vrancken for useful discussions. 
HK acknowledges support from a Clarendon Bursary.



\clearpage

FIGURE CAPTIONS

Fig. 1: Schematic diagram of a cylindrical drop.

Fig. 2: The dynamics of the advancing contact line for $a_1 / a_2 = 1$, $\theta_1=30^{\mathrm{o}}$ 
and $\theta_2=60^{\mathrm{o}}$. (a) Free energy, (b) drop radius, and (c)
macroscopic contact angle as a function of drop size. A and A$^{\prime}$: the contact line 
immediately wets the hydrophilic stripe. AB: the contact line is pinned until $\theta=\theta_2$. 
BA$^{\prime}$: the contact line advances with $\theta=\theta_2$. The inset in (a) summarises the evolution of 
the drop shape. The solid lines in the inset show the drop profiles at A, B and A$^{\prime}$, 
while the dashed lines correspond to the drop profiles at A and A$^{\prime}$ immediately before 
contact line jumps occur.

Fig. 3: The dynamics of the receding contact line for $a_1 / a_2 = 1$, $\theta_1=30^{\mathrm{o}}$ 
and $\theta_2=60^{\mathrm{o}}$. (a) Free energy, (b) drop radius, and (c)
macroscopic contact angle as a function of drop size. A and A$^{\prime}$: the contact line 
immediately dewets the hydrophobic stripe. AB: the contact line is pinned until $\theta=\theta_1$. 
BA$^{\prime}$: the contact line recedes with $\theta=\theta_1$. The inset in (a) summarises the evolution of 
the drop shape. The solid lines in the inset show the drop profiles at A, B and A$^{\prime}$, while 
the dashed lines correspond to the drop profiles at A and A$^{\prime}$ immediately before 
contact line jumps occur.

Fig. 4: Hysteresis curve for a two dimensional, cylindrical drop on a chemically striped surface.
$a_1 / a_2 = 1$, $\theta_1=30^{\mathrm{o}}$ and $\theta_2=60^{\mathrm{o}}$.
The data points are results from the lattice Boltzmann simulations and the solid line corresponds 
to the analytical calculations.

Fig. 5: Schematic diagram of the chemically patterned surface.

Fig. 6: Advancing contact line dynamics for surface A, where squares with $\theta_e=110^{\mathrm{o}}$ and length $a=12$ are separated by a distance $b=5$, on a 
background with $\theta_e=60^{\mathrm{o}}$. (a) Volume dependence of the interface position at an angle $0^{\mathrm{o}}$, $15^{\mathrm{o}}$, $30^{\mathrm{o}}$, and 
$45^{\mathrm{o}}$ with respect to the lattice axes. (b) Macroscopic contact angle as a function of drop volume. (c) Contour plots of the base radius at various drop volumes. 
(d-g) Top view of the drops at the points indicated in frame (b). (h-i) Side views of the drop as indicated in frame (g).

Fig. 7: Advancing contact line dynamics for surface B, where squares with $\theta_e=60^{\mathrm{o}}$ and length $a=12$ are separated by a distance $b=5$, on a 
background with $\theta_e=110^{\mathrm{o}}$.
(a) Volume dependence of the interface position at an angle $0^{\mathrm{o}}$, $15^{\mathrm{o}}$, $30^{\mathrm{o}}$, and $45^{\mathrm{o}}$ with respect to the lattice axes.
(b) Macroscopic contact angle as a function of drop volume. (c) Contour plots of the base radius at various drop volumes. (d-g) Top view of the drops at the points
indicated in frame (b). (h-i) Side views of the drop as indicated in frame (g).

Fig. 8: Receding contact line dynamics for surface A, where squares with $\theta_e=110^{\mathrm{o}}$ and length $a=12$ are separated by a distance $b=5$, on a 
background with $\theta_e=60^{\mathrm{o}}$.
(a) Volume dependence of the interface position at an angle $0^{\mathrm{o}}$, $15^{\mathrm{o}}$, $30^{\mathrm{o}}$, and $45^{\mathrm{o}}$ with respect to the lattice axes.
(b) Macroscopic contact angle as a function of drop volume. (c) Contour plots of the base radius at various drop volumes. (d-g) Top view of the drops at the points
indicated in frame (b). (h-i) Side views of the drop as indicated in frame (g).

Fig. 9: Receding contact line dynamics for surface B, where squares with $\theta_e=60^{\mathrm{o}}$ and length $a=12$ are separated by a distance $b=5$, on a 
background with $\theta_e=110^{\mathrm{o}}$.
(a) Volume dependence of the interface position at an angle $0^{\mathrm{o}}$, $15^{\mathrm{o}}$, $30^{\mathrm{o}}$, and $45^{\mathrm{o}}$ with respect to the lattice axes.
(b) Macroscopic contact angle as a function of drop volume. (c) Contour plots of the base radius at various drop volumes. (d-g) Top view of the drops at the points
indicated in frame (b). (h-i) Side views of the drop as indicated in frame (g).

Fig. 10: Schematic diagram of a cylindrical drop on a topologically patterned surface: (a) suspended state
and (b) collapsed state.

Fig. 11: Drop shape as a function of time from lattice Boltzmann simulations of a cylindrical drop suspended
on a topologically patterned surface. (a) The advancing contact line remains pinned during the simulation 
($a=7$ and $b=13$). (b) A contact line jump occurs when $\theta = 155^{\mathrm{o}}$ ($a=5$ and $b=5$). 
(c) The measured receding contact angle $\sim 120^{\mathrm{o}}$ ($a=7$ and $b=13$). The position of the 
contact lines can be seen more clearly in the insets.

Fig. 12: Drop shape as a function of time from lattice Boltzmann simulations of a cylindrical drop 
collapsed on a topologically patterned surface showing the 
contact line motion as the drop volume is (a) increased or (b-c) decreased. (a) and (b): $a=5$,  $b=5$ and $l=5$. 
(c): $a=5$,  $b=35$, and $l=10$. The equilibrium contact angle $\theta_e = 120^{\mathrm{o}}$. The position of the 
contact lines can be seen more clearly in the insets.

Fig. 13: Advancing contact line dynamics for a suspended drop on a superhydrophobic surface with
$a=3$, $b=7$, $l=5$, and $\theta_e = 110^{\mathrm{o}}$. (a) Contact angle as a function of volume.
(b) and (c) Top and side view of the drop at $V=6.8 \, 10^{5}$. (d) and (e) Top and side view of 
the drop at $V=16.8 \, 10^{5}$. The insets in frame (b) and (d) are the corresponding contour plots
of the base of the drop.

Fig. 14: Receding contact line dynamics for a suspended drop on a superhydrophobic surface with
$a=3$, $b=7$, $l=5$, and $\theta_e = 110^{\mathrm{o}}$. (a) Contact angle as a function of volume.
(b) and (c) Top and side view of the drop at $V=15.4 \, 10^{5}$. (d) and (e) Top and side view of 
the drop at $V=15.1 \, 10^{5}$. The insets in frame (b) and (d) are the corresponding contour plots
of the base of the drop.

Fig. 15: Advancing contact line dynamics for a collapsed drop on a superhydrophobic surface with
$a=4$, $b=6$, $l=5$, and $\theta_e = 120^{\mathrm{o}}$. (a) Contact angle as a function of volume. (b--e)
Contour plots of the drop at positions indicated in (a): the different lines correspond to heights 1, 3, 5, and 6 lattice spacings 
away from the bottom surface. (f) The drop cross section in the horizontal direction.
(g) The drop cross section in the diagonal direction. The spherical fit is shown by the dashed lines
and the inset magnifies the spherical fit close to the posts.

Fig. 16: Receding contact line dynamics for a collapsed drop on a superhydrophobic surface with
$a=4$, $b=6$, $l=5$, and $\theta_e = 120^{\mathrm{o}}$. (a) Contact angle as a function of volume. (b--e)
Contour plots of the drop at positions indicated in (a): the different lines correspond to heights 1, 3, 5, and 6 
lattice spacings away from the bottom surface.


\clearpage

\clearpage

\clearpage

\clearpage

\clearpage

\clearpage

\clearpage

\clearpage

\clearpage

\clearpage

\clearpage

\clearpage

\clearpage

\clearpage

\clearpage

\clearpage


\end{document}